%% file: draft.tex
\documentclass[journal,10pt]{IEEEtran}
\usepackage[top=0.67in,bottom=0.99in,left=0.67in,right=0.67in]{geometry}
\usepackage[english]{babel}
\usepackage{graphicx}
\usepackage{subfig}
\usepackage{multicol}
\usepackage{setspace}
\usepackage{amsmath}
\usepackage{amsthm}
\usepackage{mathtools}
\usepackage{epstopdf}
\usepackage{fancyhdr}
\usepackage{indentfirst}
\usepackage{enumerate}
\usepackage{caption}
\usepackage{leftidx}
\usepackage{amssymb}
\usepackage{float}
\usepackage{breqn}
\usepackage{color, xcolor}
\usepackage{bm}
\usepackage{cite}
\usepackage{multirow}
\usepackage{totcount}
\usepackage{algorithm}
\usepackage{algcompatible}
\usepackage{hyperref}
\usepackage{bbm}
\usepackage{adjustbox}
\usepackage{array}
\usepackage{acro}
\usepackage{chngcntr}
\usepackage{booktabs}
\usepackage{multirow}
\usepackage{makecell}

\counterwithin*{footnote}{page}

\hypersetup{nolinks=true}

\newtheoremstyle{colon}%
{}%Space above
{}%Space below
{\itshape}%Bodyfont
{}%Indent amount (empty = no indent, \parindent = para indent)
{\bfseries}%Headfont
{:\ }%Punctuation after thm head
{ }%Space after thm head: " " = normal interword space;\newline = linebreak
{\thmname{#1}\thmnumber{\bfseries\ #2}\thmnote{\ (#3)}}%Thm head spec (can be left empty, meaning `normal')

\theoremstyle{colon}

\newtheorem{Thm}{Theorem}[section] % Theorem numbering
\renewcommand{\theThm}{\arabic{section}.\arabic{Thm}} % Theorem numbering format
\newtheorem{Def}[Thm]{Definition} % Definition numbering aligned with Theorem

\newtheorem{Lem}[Thm]{Lemma} % Lemma numbering aligned with Theorem

\newtheorem{Obs}[Thm]{Observation} % Observation numbering aligned with Theorem

\newcommand{\bn}{\textnormal}

\IEEEaftertitletext{\vspace{-1.5\baselineskip}}
\input{acronyms.tex} 
\begin{document}
\title{\huge 3D Near-Field Beam Training for Uniform Planar Arrays through Beam Diverging}

\author{
\thanks{This work has been partially accepted for presentation at the 2025 IEEE Global Communications Conference (GLOBECOM). ({\itshape Corresponding author: Ying-Jun Angela Zhang.})}
\IEEEauthorblockN{Ran~Li,~\textit{Member,~IEEE}\vspace{-0.1cm}, Ziyi~Xu,~\textit{Graduate Student Member,~IEEE,} and Ying-Jun~Angela~Zhang,~\textit{Fellow,~IEEE}\vspace{-0.3cm}}
\thanks{R. Li, Z. Xu, and Y.-J. A. Zhang are with the Department of Information Engineering, The Chinese University of Hong Kong, Hong Kong SAR (e-mails: \{ranli,xz022,yjzhang\}@ie.cuhk.edu.hk).}
}

\maketitle
\thispagestyle{empty}

\begin{abstract}
% Background
In future 6G communication systems, large-scale antenna arrays promise enhanced signal strength and spatial resolution, but they also increase the complexity of beam training.
Moreover, as antenna counts grow and carrier wavelengths shrink, the channel model transits from far-field (FF) planar waves to near-field (NF) spherical waves, further complicating the beam training process.
% Our mean scope
This paper focuses on millimeter-wave (mmWave) systems equipped with large-scale uniform planar arrays (UPAs), which produce 3D beam patterns and introduce additional challenges for NF beam training.
% Challenge
Existing methods primarily rely on either FF steering or NF focusing codewords, both of which are highly sensitive to mismatches in user equipment (UE) location, leading to high sensitivity to even slight mismatch and excessive training overhead.
% Our method
In contrast, we introduce a novel beam training approach leveraging the \textbf{beam-diverging effect}, which enables adjustable wide-beam coverage using only a single radio frequency (RF) chain. Specifically, we first analyze the spatial characteristics of this effect in UPA systems and leverage them to construct hierarchical codebooks for coarse UE localization. Then, we develop a 3D sampling mechanism to build an NF refinement codebook for precise beam training.
% Results
Numerical results demonstrate that the proposed algorithm achieves superior beam training performance while maintaining low training overhead.

\end{abstract}

\spacing{0.98}

\section{Introduction}
With the emergence of data-intensive applications such as digital twins and augmented reality, the spectral efficiency requirements of 6G networks are projected to increase by an order of magnitude \cite{RoadTo6G}. To meet this demand for significantly higher data rates, future wireless systems are increasingly turning to the \ac{mmWave} frequency band and large-scale antenna arrays as critical enabling technologies \cite{ELAAMIMO}.
On the one hand, large antenna arrays at the transmitter allow \ac{mmWave} systems to perform highly directional beamforming, which helps maintain a robust and focused received signal power. However, this narrow-beam operation comes with a tradeoff: even slight misalignment between the beam direction and the target can cause a sharp drop in received signal strength, severely degrading overall system performance.
Consequently, accurate beam alignment between the transmitter and receiver is vital for reliable mmWave communication \cite{SAMBA}, underscoring the need for efficient and low-overhead beam training strategies.

On the other hand, as the \ac{BS} array size increases and the wavelength shortens, \acp{UE} are more likely to fall within the \ac{NF} region of the \ac{BS} \cite{TutReview}. In this regime, the spherical nature of the wavefront enables the concentration of power at specific spatial locations rather than along angular directions, introducing strong dependencies on both angle and distance, significantly complicates the beam training process.

In this work, we focus on near-field beam training for \ac{UPA} where antenna elements are evenly spaced across a plane.
Unlike \acp{ULA}, which arrange elements along a single axis, \acp{UPA} enable compact deployment of large antenna arrays within a limited footprint.
The planar structure allows \acp{UPA} to generate 3D beamforming patterns, in contrast to the 2D patterns produced by \acp{ULA} \cite{UPAregion}.
The utilization of \acp{UPA} is well recognized in both industry and standards. Their use is supported by the 3GPP Release 18 specifications \cite{3GPPR18} and has been widely adopted in modern cellular systems, particularly for 5G and beyond \cite{UPAcell}. For instance, small cells operating in the mmWave band have been extensively deployed in dense urban environments, where UPAs are essential for precise coverage control, minimizing signal leakage, and mitigating co-channel interference from neighboring cells.
Moreover, UPAs are particularly advantageous in multi-altitude communication scenarios. They can simultaneously serve users across different floors of high-rise buildings and facilitate robust links between unmanned aerial vehicles (UAVs) and ground users, thereby supporting flexible and reliable aerial-ground communication \cite{UAVandUPA}.

% Given the above advantages and floushishment of \acp{UPA}, in this paper, we study the \ac{UPA} beam training technique, especially in the near-field.

\subsection{Existing Works}
Beam training has been extensively investigated across different array configurations and propagation regimes. However, limited attention has been given to the dimensionality challenge inherent in near-field beam training with \acp{UPA}. In conventional settings, the resulting codebook size grows rapidly with the number of antennas, leading to prohibitively high search complexity and storage requirements.
Existing methods for beam training in this context can be broadly categorized into the following approaches:

\textbf{Far-Field Beam Training for \acp{ULA} and \acp{UPA}}.
Conventional \ac{FF} beam training typically performs exhaustive search over a predefined codebook \cite{3GPPR16}, resulting in high latency and energy consumption.
To mitigate this, hierarchical beam search methods\cite{Hierarchical} use multi-tier codebooks, starting with wide beams and progressively refining the search.
In parallel, data-driven approaches have emerged, where neural networks predict the best beam index from a ULA codebook \cite{LSSP, SAMBA}.

While \ac{ULA}-based methods are well-understood and widely used \cite{3GPPR16, Hierarchical, SAMBA}, extending them to UPAs is more complex due to the need for joint azimuth-elevation beamforming, which greatly expands the codebook size. A common simplification is to decouple the UPA into horizontal and vertical ULAs \cite{partition}. Alternatively, learning-based methods like \cite{GridFree} bypass codebooks by treating beamforming as a regression problem, generating continuous beamformers instead of selecting from discrete candidates.

\textbf{Near-Field Beam Training for \acp{ULA}}.
Recent studies have extended beam training to the \ac{NF} regime, where spherical wavefronts and distance-dependent focusing effect are critical. For \acp{ULA}, polar-domain codebooks have been proposed to steer energy towards specific spatial locations \cite{DaiChannelEstimation}. Fresnel transform-based codebook is also introduced in \cite{DFnT} to reduce codeword interference.

As in the \ac{FF}, many \ac{NF} schemes adopt multi-stage search with progressively refined resolution. For instance, \cite{YouBeamTraining} proposes a two-tier method: the \ac{BS} first performs coarse angular estimation using \ac{FF} beams, then refines the result using NF polar beams. Similarly, \cite{SpatialChirp} employs multi-tier codebooks to progressively narrow the search space, though spatial overlaps among beams may arise due to imperfect beam patterns.
To improve coverage efficiency, \cite{GratingLobes} utilized grating lobes to extend spatial coverage with a single RF chain. \cite{SuperResolutionYou} further designed beamformers with adjustable beamwidths to support multi-resolution search, though this incurs high computational due to the beamformer optimization and requires amplitude control using multiple RF chains.
Beyond model-driven approaches, deep learning has been applied to \ac{NF} \acp{ULA} beam training, where neural networks approximate the complex beam selection function \cite{SenseThenTrain, NFbeamTrainingDL}.
% In \cite{SenseThenTrain}, a DNN maps received ping-pong pilots at the BS to appropriate narrow beams. Similarly, \cite{NFbeamTrainingDL} uses downlink NF pilots and UE feedback to jointly predict angle and distance using a \ac{DNN}.

\begin{figure}[t]
    \centering
    \includegraphics[width=0.87\columnwidth]{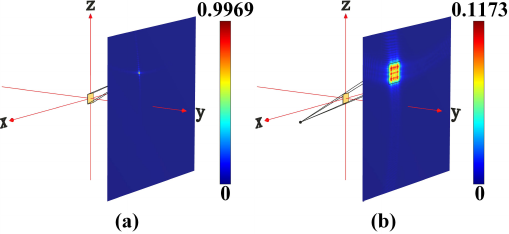}
    \captionsetup{font={small}}
    \caption{Amplitude of the received pilot on a plane orthogonal to the $y$-axis. The orange block at the origin denotes the UPA. (a) \textbf{Beam-focusing} codeword yields a concentrated pattern. (b) \textbf{Beam-diverging} codeword (defined later) provides broader coverage.}\label{fig:I_focus_diverge}
    \vspace{-0.2cm}
\end{figure}

\textbf{Near-Field Beam Training for \acp{UPA}.}
\ac{NF} beam training for \acp{UPA} remains relatively underexplored due to its intrinsic complexity. Early studies \cite{NFris, NearFieldHier, NFchannelEstimationUPA, UPAaperture, DFTforNF} have emphasized the challenges arising from the large-scale antenna array and the distance-dependent focusing effect. Recent efforts have primarily aimed to reduce training complexity \cite{NearFieldHier, DFTforNF}.
For example, \cite{DFTforNF} proposed a subarray partitioning strategy that applies conventional far-field DFT codebooks to individual UPA subarrays, enabling hierarchical training. While this design simplifies codebook generation, it requires multiple RF chains to support simultaneous subarray probing.

Despite these advances, a fundamental challenge persists across both near- and far-field regimes, and for both linear and planar arrays: how to efficiently generate beams with controllable resolution while keeping complexity and hardware cost low. This issue is especially critical for NF UPA beam training, which demands scalable solutions capable of navigating high-dimensional search spaces without compromising hardware efficiency—a problem that remains largely open.

\subsection{Limitations of Existing Approaches and Our Contributions}
To put it in detail, the challenges of \ac{NF} \ac{UPA} beam training come from multiple aspects, which include:

\begin{itemize}
    \item \textit{\textbf{Large codebook size}}: UPAs typically contain massive antennas elements due to their 2D structure, generally leading to numerous beam training codewords \cite{NFris, NFchannelEstimationUPA}. This complexity is further amplified in the NF, where the 3D nature of the channel, including azimuth, elevation, and distance, greatly expands the codebook size and search space.
    \item \textit{\textbf{Lack of efficient hierarchical search method}}:
    While hierarchical schemes \cite{Hierarchical, NearFieldHier} provide accurate and low-complexity solutions in the FF, their effectiveness diminishes in the NF. The focusing beam not only is extremely misalignment-sensitive, but also exhibits strong interference across adjacent distance rings \cite{TutReview}, potentially causing misjudgment during the search. 
    On one hand. this sharp focusing beam hinders the effective capture of user's existence, especially in the initial coarse search stage.
    On the other hand, severe leakage in radiation power significantly affect the development of efficient multi-tier strategies. 
    
\end{itemize}

To address the challenges of NF beam training for \ac{UPA} systems, we propose a 3D beam training scheme based on the beam diverging effect. In our prior work, \cite{li2025icc}, we explored this phenomenon as a counterpart to the well-studied NF focusing effect \cite{TutReview, DaiChannelEstimation, YouBeamTraining, NFris}, which is analogous to light focusing through a convex lens, as shown in Fig. \ref{fig:I_focus_diverge}. By contrast, the beam diverging effect resembles light diverging and is more suitable for generating wide, controllable beams that facilitate efficient beam training.
Specifically, this work makes the following contributions:

\begin{itemize}
    \item \textbf{Beam-Diverging Codewords for 3D Coverage:} 
    Inspired by the behavior of concave lenses—which diverge light rays away from the optical axis—we analyze the beam divergence effect in a 2D source setting, as illustrated on the left side of Fig.~\ref{fig:I_focus_diverge}. Our focus is on designing diverging beams with broad spatial coverage, making them naturally suited for multi-tier beam training, particularly in \ac{NF} UPA scenarios.
    We show that diverging beams generated by UPAs can achieve seamless 3D spatial coverage. Moreover, their beamwidth can be flexibly controlled via analog beamforming, enabling efficient coverage with a single RF chain and thus providing a hardware-friendly solution.
    \item \textbf{Hierarchical Codebook Design:}
    We propose a multi-resolution codebook that partitions the 3D space into hierarchical four-sided pyramidal regions. This structure naturally enables multi-tier beam search, substantially reducing the training overhead.
    In contrast, existing methods often rely on exhaustive grid-based searches \cite{NFris}, which are costly in terms of latency and energy, or require complex amplitude control techniques \cite{NearFieldHier, UPAregion} that depend on multiple RF chains.
    \item \textbf{Practical Beam Training Algorithm:}
    We develop a near-field UPA beam training algorithm that achieves high alignment accuracy with minimal overhead. To the best of our knowledge, it offers one of the lowest beam sweeping costs reported for NF UPA beam training.
\end{itemize}

The remainder of this article is organized as follows:
Section II presents the system model and problem formulation.
Section III analyzes the \ac{UPA} beam diverging effect.
Section IV designs the MPC for beam training, while Section V proposes techniques to enhance its performance.
Section VI provides simulation results, and Section VII concludes the paper.

\section{System Model and Problem Formulation}

\subsection{UPA Model}

\begin{figure}[t]
    \centering
    \includegraphics[width=0.8\columnwidth]{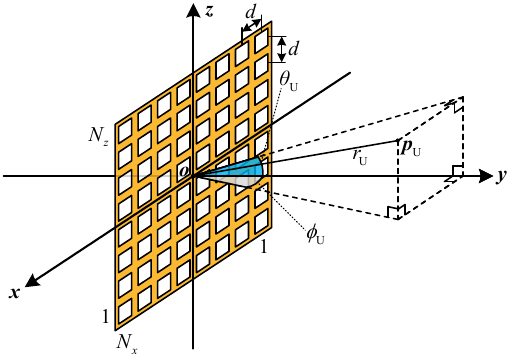}
    \captionsetup{font={small}}
    \caption{UPA centered at the origin $\bm{o}$ and UE at $\bm{p}_{\bn{U}}$.}\label{fig:II_system}
\end{figure}

As illustrated in Fig. \ref{fig:II_system}, we consider a wireless system where the BS uses a UPA to communicate with a single-antenna UE. The UPA is oriented orthogonally to the $y$-axis, with its center located at the origin $\bm{o}\triangleq(0, 0, 0)$. Meanwhile, it consists of $N_x$ antennas per row along the $x$-axis and $N_z$ antennas per column along the $z$-axis, with an antenna spacing of $d\triangleq\frac{\lambda}{2}=\frac{c}{2f}$, where $\lambda$ and $f$ are the carrier wavelength and frequency, respectively. Consequently, the UPA has a rectangular aperture with side lengths of $D_x\triangleq(N_x-1)d$ and $D_z\triangleq(N_z-1)d$. The Cartesian coordinates of the $(x,z)$-th UPA antenna are derived as $\bm{p}_{x,z}\triangleq(\frac{2x-N_x-1}{2}d,0,\frac{2z-N_z-1}{2}d)$ with $x\in\mathcal{N}_x\triangleq\{1,2,\cdots,N_x\}$ and $z\in\mathcal{N}_z\triangleq\{1,2,\cdots,N_z\}$. The Cartesian coordinates of the UE are denoted as $\bm{p}_{\bn{U}}\triangleq(x_{\bn{U}}, y_{\bn{U}}, z_{\bn{U}})$ and are unknown to the BS. We also denote the Euclidean distance between the origin $\bm{o}$ and $\bm{p}_{\bn{U}}$ as $r_{\bn{U}}\triangleq\|\overrightarrow{\bm{op}_{\bn{U}}}\|$, the angle between the projection of the vector $\overrightarrow{\bm{op}_{\bn{U}}}$ onto the $x$-$y$ plane and the positive $y$-axis as $\phi_{\bn{U}}$, and the angle between its projection onto the $y$-$z$ plane and the positive $y$-axis as $\theta_{\bn{U}}$. Apparently, it follows that $x_{\bn{U}}=y_{\bn{U}}\tan\phi_{\bn{U}}$ and $z_{\bn{U}}=y_{\bn{U}}\tan\theta_{\bn{U}}$\footnote{In this paper, the UPA communicates only with UEs located within an unbounded rectangular pyramidal frustum defined by $\mathcal{R}\!\triangleq\!\{\bm{p}_{\bn{U}}|-y_{\bn{U}}\!-\!\frac{D_x}{2}$ $<\!x_{\bn{U}}<\!y_{\bn{U}}\!+\!\frac{D_x}{2},\!y_{\bn{U}}\!>\!0,\!-y_{\bn{U}}\!-\!\frac{D_z}{2}\!<\!z_{\bn{U}}\!<\!y_{\bn{U}}\!+\!\frac{D_z}{2}\}$, which can be approximately expressed as $\mathcal{R}\!\approx\!\{\bm{p}_{\bn{U}}|y_{\bn{U}}\!>\!0,-\frac{\pi}{4}\!<\!\phi_{\bn{U}}\!<\!\frac{\pi}{4},-\frac{\pi}{4}\!<\!\theta_{\bn{U}}\!<\!\frac{\pi}{4}\}$. Under this configuration, a BS above the ground can communicate with UEs at any location using six UPAs, each oriented toward one of the six directions: $\pm x$, $\pm y$, and $\pm z$. An additional configuration, where the UPA communicates with UEs within a larger frustum $\mathcal{R}=\{\bm{p}_{\bn{U}}|y_{\bn{U}}\!>\!0,-\frac{\pi}{3}\!<\!\phi_{\bn{U}}\!<\!\frac{\pi}{3},-\frac{\pi}{3}\!<\!\theta_{\bn{U}}\!<\!\frac{\pi}{3}\}$, is discussed in Section \ref{sec:num}.}.

\subsection{NF Region for UPA}

The NF region refers to the area where assuming the signal propagates as a planar wave leads to significant beamforming degradation, and thus the spherical wave propagation model is used instead. In general, the NF region for UPA is approximated as a hemisphere, defined by $\mathcal{R}_{\bn{hem}}\triangleq\{\bm{p}_{\bn{U}}|y_{\bn{U}}>0,r_{\bn{U}}\leq\frac{2D^2}{\lambda}\}$, where the radius $\frac{2D^2}{\lambda}$ is the Rayleigh distance and $D\triangleq\sqrt{D_x^2+D_z^2}$ is the diagonal of the UPA's aperture, as described in \cite[Chapter 2.2.4]{balanis2016antenna}. Here, we further demonstrate that the NF region for UPA is equivalent to the union of the NF regions of two ULAs aligned along the diagonals of the UPA. An illustrative example is shown in Fig. \ref{fig:II_nf}, and the rigorous proof is provided in Appendix \ref{app:I}. It can also be validated that the NF region for UPA can occupy up to $54\%$ of the volume of the aforementioned hemisphere $\mathcal{R}_{\bn{hem}}$, underscoring the necessity of designing NF-specific methods to effectively serve UEs near the BS.
\begin{figure}[t]
\vspace{-0.4cm}
    \centering
    \includegraphics[width=\columnwidth]{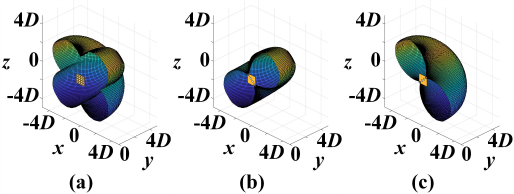}
    \captionsetup{font={small}}
    \caption{Example illustrating the NF region for a UPA with $N_x=N_z=4$ and $f=28$ GHz. (a) NF region boundary for UPA; (b) NF region boundary for a ULA with 4 antennas aligned along $x=-z$; (c) NF region boundary for a ULA with 4 antennas aligned along $x=z$.}\label{fig:II_nf}
    \vspace{-0.2cm}
\end{figure}
\begin{figure*}[t]
    \centering
    \includegraphics[width=7in]{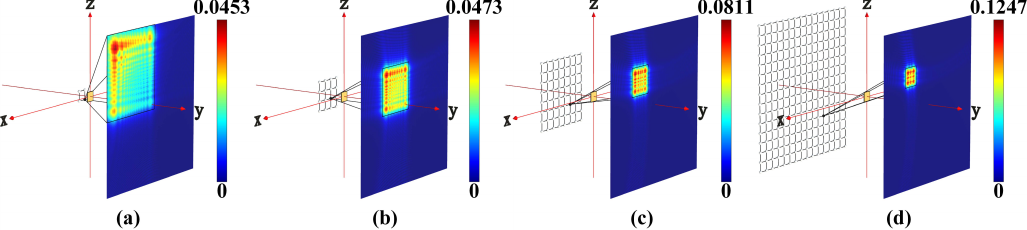}
    \captionsetup{font={small}}
    \caption{Amplitude of the received pilot on $\mathcal{R}_{3D}$ (i.e, the UE plane sampled at $y_{\bn{U}}=3D$) when deploying $\bm{c}(\bm{v})$ as the pilot. The black sphere denotes the virtual focal point $\bm{v}$ and the black lines on the sampled UE plane represent the boundary of $\mathcal{R}_{3D}(\bm{v})$. The UPA is configured with $N_x=N_z=128$ and $f=28$ GHz. Thus, $D_x=D_z$ holds. Subfigures correspond to scenarios with different virtual focal points: (a) $\bm{v}=(-D_x/2,-D_x,-D_z/2)$; (b) $\bm{v}=(-D_x/2,-2D_x,-D_z/2)$; (c) $\bm{v}=(-3D_x/2,-4D_x,-3D_z/2)$; (d) $\bm{v}=(-7D_x/2,-8D_x,-7D_z/2)$.}\label{fig:III_method1_1}
    \vspace{-0.2cm}
\end{figure*}
\subsection{Problem Formulation}
This paper focuses on the downlink scenario where the BS sequentially transmits $T$ pilots to a UE located within the NF region for beam training. Specifically, we assume that the downlink channel follows a Rician fading model, comprising one LOS path and $L$ NLOS paths, formulated as 
\begin{align*}
\bm{h}\!\triangleq\sum_{l=0}^Lg_l\bm{b}(\bm{s}_l).
\end{align*}
Here, $\bm{s}_0=\bm{p}_{\bn{U}}$ and $\bm{s}_l$ (for $l>0$) denote the Cartesian coordinates of the UE and the scatter associated with the $l$-th NLOS path, respectively. $g_0$ and $g_l$ (for $l>0$) denote the complex channel gains of the LOS and $l$-th NLOS paths, respectively. $\bm{b}(\bm{s}_l)$ denotes the NF steering vector and is defined as  
\begin{align*}
\begin{split}
\bm{b}(\bm{s}_l)\!\triangleq\!\Big[&e^{-j\frac{2\pi}{\lambda}\|\overrightarrow{\bm{p}_{1,1}\bm{s}_l}\|}\!,e^{-j\frac{2\pi}{\lambda}\|\overrightarrow{\bm{p}_{1,2}\bm{s}_l}\|}\!,\cdots,e^{-j\frac{2\pi}{\lambda}\|\overrightarrow{\bm{p}_{1,N_z}\bm{s}_l}\|},\\
&\qquad\qquad\qquad\quad\vdots\\
&e^{-j\frac{2\pi}{\lambda}\|\overrightarrow{\bm{p}_{N_x\!,1}\!\bm{s}_l}\|}\!,e^{-j\frac{2\pi}{\lambda}\|\overrightarrow{\bm{p}_{N_x\!,2}\!\bm{s}_l}\|}\!,\cdots,e^{-j\!\frac{2\pi}{\lambda}\|\overrightarrow{\bm{p}_{N_x\!,N_z}\!\bm{s}_l}\|}\!\Big]^{T}\!\!\!.
\end{split}
\end{align*}

Let $\bm{w}(t)\in\mathbb{C}^{N_xN_z\times 1}$ denote the $t$-th pilot transmitted by the BS, where $\mathbb{C}$ is the set of complex numbers. Then, the $t$-th received pilot at the UE can be expressed as $y(t)\triangleq\bm{h}^T\bm{w}(t)+n(t)$, where $n(t)$ is the complex Gaussian noise distributed as $\mathcal{CN}(0,\sigma^2)$. Similarly, $T$ received pilots can be collectively expressed as 
\begin{align*}
\bm{y}^T=\bm{h}^T\bm{W}+\bm{n}^T,
\end{align*}
where $\bm{y}\!\triangleq\![y(1),y(2),\cdots,y(T)]^T\!\!$, $\bm{W}\!\triangleq\![\bm{w}(1),\bm{w}(2),\cdots,$ $\bm{w}(T)]$, and $\bm{n}\triangleq[n(1),n(2),\cdots,n(T)]^T$ follows distribution $\mathcal{CN}(0,\sigma^2\bm{I}_T)$.

The goal of beam training is to identify the codeword that is optimally matched to the channel while using as few pilots as possible. In this paper, we adopt the signal-to-noise ratio (SNR) as the performance metric and formulate the beam training problem as 
\begin{align*}
\max_{\bm{w}}\ \frac{|\bm{h}^T\bm{w}|^2}{\sigma^2}, 
\end{align*}
where the pilot overhead $T$ associated with $\bm{W}$ is also expected to be minimized.
\section{Beam Diverging Effect for UPA}
In this section, we first define the diverging codewords for UPA and illustrate how they induce the beam diverging effect. Next, we provide an intuitive interpretation to clarify the rationale behind this effect. Finally, we introduce the identification accuracy as a metric to characterize the beam diverging effect, thereby demonstrating its effectiveness in facilitating efficient beam training.
\subsection{Diverging Codewords and Beam Diverging Effect for UPA}
The diverging codewords for UPA are defined as follows:
\begin{Def}[Diverging Codeword for UPA]\label{def:mword}
Given any point $\bm{v}=(x_{\bm{v}},y_{\bm{v}},z_{\bm{v}})$ satisfying $y_{\bm{v}}<0$, we define its diverging codeword as
\begin{align}\label{eq:mword}
\bm{c}(\bm{v})\triangleq\frac{1}{\sqrt{\!N_xN_z}}\bm{b}(\bm{v})
\end{align}
\end{Def}
Similar to the case of ULA in \cite{li2025icc}, we refer to $\bm{c}(\bm{v})$ as the diverging codeword since it not only exhibits opposite phases relative to the focusing codeword—as can be verified, the focusing codeword at point $\bm{v}$ is $\overline{\bm{c}(\bm{v})}$—but also induces a diverging beam pattern. To illustrate this, we deploy the codeword $\bm{c}(\bm{v})$ as the pilot and plot the (normalized) amplitude of the received pilot, i.e., $|\frac{1}{\sqrt{N_xN_z\sum_{l=0}^Lg_l^2}}\bm{h}^T\bm{c}(\bm{v})|\approx|\bm{c}(\bm{p}_{\bn{U}})^T\bm{c}(\bm{v})|\in[0,1]$\footnote{This approximation intentionally neglects the NLOS component in $\bm{h}$, since the beam diverging effect discussed below is induced solely by the LOS component. Although the NLOS components also always present, they typically have low power in the NF regime and thus do not significantly influence the presence or characteristics of the effect.}, in Fig. \ref{fig:III_method1_1}. The results are formally stated below.
\begin{Obs}[Beam Diverging Effect for UPA]\label{obs:bde}
When deploying the diverging codeword $\bm{c}(\bm{v})$ as the pilot, the normalized power of the received pilot (i.e., the square of the normalized amplitude) is substantially higher when the UE lies on the rectangular planar region 
\begin{align}
\begin{split}\label{def:rect}
\!\!\mathcal{R}_{\hat{y}}\!(\!\bm{v}\!)\!\!\triangleq\!\!\Big\{\!\bm{p}_{\bn{U}}\!\!\in\!\!\mathcal{R}\!\Big|\!&\Big(\!x_{\bm{v}}\!\!+\!\!\frac{D_x}{2}\!\!\Big)\frac{y_{\bn{U}}}{y_{\bm{v}}}\!\!-\!\!\frac{D_x}{2}\!\!<\!\!x_{\bn{U}}\!\!<\!\!\Big(\!x_{\bm{v}}\!\!-\!\!\frac{D_x}{2}\!\!\Big)\frac{y_{\bn{U}}}{y_{\bm{v}}}\!\!+\!\!\frac{D_x}{2}\!,\\
&y_{\bn{U}}=\hat{y},\\
&\Big(\!z_{\bm{v}}\!\!+\!\!\frac{D_z}{2}\!\!\Big)\frac{y_{\bn{U}}}{y_{\bm{v}}}\!\!-\!\!\frac{D_z}{2}\!\!<\!\!z_{\bn{U}}\!\!<\!\!\Big(\!z_{\bm{v}}\!\!-\!\!\frac{D_z}{2}\!\!\Big)\frac{y_{\bn{U}}}{y_{\bm{v}}}\!\!+\!\!\frac{D_z}{2}\!\Big\}\!,
\end{split}
\end{align}
compared to when the UE lies on the complement of $\mathcal{R}_{\hat{y}}(\bm{v})$, defined as $\bar{\mathcal{R}}_{\hat{y}}(\bm{v})\triangleq\mathcal{R}_{\hat{y}}\setminus\mathcal{R}_{\hat{y}}(\bm{v})$. Here, $\hat{y}$ can be any positive number no smaller than the Fresnel distance\footnote{Fresnel distance is generally small and given by $0.62\sqrt{D^3/\lambda}$ (see \cite[Chapter 2.2.4]{balanis2016antenna}). The region within the Fresnel distance, i.e., $\{\bm{p}_{\bn{U}}\in\mathcal{R}|y_{\bn{U}}<0.62\sqrt{D^3/\lambda}\}$, is dominated by the reactive field and is therefore excluded from our analysis.}, and $\mathcal{R}_{\hat{y}}\!\triangleq\!\{\bm{p}_{\bn{U}}\!\in\mathcal{R}|y_{\bn{U}}=\hat{y}\}$ denotes the UE plane sampled at $y_{\bn{U}}=\hat{y}$.
\end{Obs}

\begin{figure}[t]
    \centering
    \includegraphics[width=0.8\columnwidth]{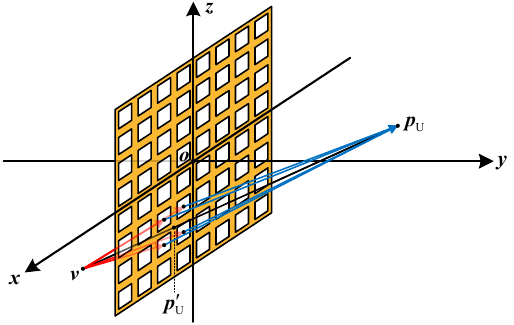}
    \captionsetup{font={small}}
    \caption{Example demonstrating beam diverging effect for UPA.}\label{fig:II_geo}
    \vspace{-0.2cm}
\end{figure}

Remarkably, $\mathcal{R}_{\hat{y}}(\bm{v})$ is exactly the geometric projection of the UPA's aperture onto $\mathcal{R}_{\hat{y}}$ along rays emitted from $\bm{v}$. Therefore, the beam diverging effect for UPA is analogous to the optical diverging effect produced by a concave lens, with the UPA acting as the lens and $\bm{v}$ as its optical virtual focal point. Based on this analogy, we refer to $\bm{v}$ as the \textbf{virtual focal point} throughout this paper. Furthermore, this diverging effect also contrasts with the optical focusing effect of a convex lens, where the UPA also acts as the lens and the focusing beam from its aperture converges toward the focusing point, analogous to how light rays converge at the optical focal point of the lens.

\subsection{Interpretation of Beam Diverging Effect for UPA}
We now provide an intuitive interpretation of why the beam diverging effect for UPA holds. Specifically, when the diverging codeword $\bm{c}(\bm{v})$ is deployed as the pilot, the amplitude of the received pilot at the UE can be calculated as
\begin{align}\label{eq:received}
|\bm{c}(\bm{p}_{\bn{U}})^T\!\bm{c}(\bm{v})|=\frac{1}{N\!_x\!N\!_z}\!\Bigg|\!\sum_{x\in\mathcal{N}_x\!,z\in\mathcal{N}_z}\!\!e^{-\!j\!\frac{2\pi}{\lambda}\!(\|\overrightarrow{\bm{vp}_{x,z}}\|+\|\overrightarrow{\bm{p}_{x,z}\bm{p}_{\bn{U}}}\|)}\!\Bigg|.	
\end{align}
If the UE location satisfies $\bm{p}_{\bn{U}}\in\mathcal{R}_{\hat{y}}(\bm{v})$, it can be proved that the line segment $\bm{vp}_{\bn{U}}$ intersects the UPA's aperture at some point. Then, for antennas located near this intersection point---say, the $(x,z)$-th antenna---the value of $(\|\overrightarrow{\bm{vp}_{x,z}}\|+\|\overrightarrow{\bm{p}_{x,z}\bm{p}_{\bn{U}}}\|)$ closely approximates $\|\overrightarrow{\bm{vp}_{\bn{U}}}\|$. As a result, the exponential terms $e^{-j\frac{2\pi}{\lambda}(\|\overrightarrow{\bm{vp}_{x,z}}\|+\|\overrightarrow{\bm{p}_{x,z}\bm{p}_{\bn{U}}}\|)}$ associated with these nearby antennas are approximately aligned in phase, which leads to constructive superposition in \eqref{eq:received} and thus a high overall received pilot amplitude. However, if the UE lies on the complement region, i.e., $\bm{p}_{\bn{U}}\in\bar{\mathcal{R}}_{\hat{y}}(\bm{v})$, the line segment $\bm{vp}_{\bn{U}}$ does not intersect with the UPA's aperture. Consequently, the values of $(\|\overrightarrow{\bm{vp}_{x,z}}\|+\|\overrightarrow{\bm{p}_{x,z}\bm{p}_{\bn{U}}}\|)$ vary significantly across different antennas, and the corresponding exponential terms are no longer phase-aligned, which leads to a relatively low overall received pilot amplitude.

\subsection{Characterization of Beam Diverging Effect for UPA}\label{sec:iden}
The beam diverging effect for UPA (or diverging codewords) can be characterized from various perspectives, such as energy leakage, misalignment sensitivity, and diverging degree \cite{li2025icc}. In this paper, we focus specifically on its effectiveness in the context of beam training and introduce the metric of identification accuracy as follows: 
\begin{Def}[Identification Accuracy for UPA]\label{def:iden}
Suppose the UE lies on the sampled UE plane $\mathcal{R}_{\hat{y}}$. Then, for any set of virtual focal points $\mathcal{V}\triangleq\{\bm{v}_n\}_{n=1}^{N}$ satisfying
\begin{align}\label{eq:accuracy}
\cup_{n=1}^N\mathcal{R}_{\hat{y}}(\bm{v}_n)=\mathcal{R}_{\hat{y}},
\end{align}
we can sequentially transmit diverging codewords $\{\bm{c}(\bm{v}_n)\}_{n=1}^N$ as pilots, and identify which $\mathcal{R}_{\hat{y}}(\bm{v}_n)$ with $n\in\{1,2,\cdots,N\}$ contains the UE by selecting the one corresponding to the highest received pilot power. The identification accuracy $\epsilon_{\hat{y}}(\mathcal{V},\gamma)$ is defined as the probability of correct identification when the reference signal-to-noise ratio (SNR) in the LOS-only regime, given by $\frac{|g_0|^2}{\sigma^2}$, equals $\gamma$. 
\end{Def}
\begin{table}[t]
\vspace{0.1cm}
\centering
\renewcommand{\arraystretch}{1}
{\footnotesize
\begin{tabular}{|>{\centering\arraybackslash}p{0.8cm}|>{\centering\arraybackslash}p{0.8cm}|>{\centering\arraybackslash}p{0.8cm}|>{\centering\arraybackslash}p{0.42cm}|>{\centering\arraybackslash}p{0.42cm}|>{\centering\arraybackslash}p{0.42cm}|>{\centering\arraybackslash}p{0.42cm}|>{\centering\arraybackslash}p{0.42cm}|>{\centering\arraybackslash}p{0.42cm}|}
\hline
\multicolumn{3}{|c|}{$\mathcal{V}\!\!=\!\!\{\!(\!\pm\!\frac{k\!_x}{2}\!\!D\!_x\!,\!-\!k\!_y\!D\!_x\!,\!\pm\!\frac{k\!_z}{2}\!\!D\!_z\!)\!\}$}&\multicolumn{2}{c|}{$k_{\hat{y}}$=0}&\multicolumn{2}{c|}{$k_{\hat{y}}$=0.3}&\multicolumn{2}{c|}{$k_{\hat{y}}$=0.7}\\
\hline
$k_x$&$k_y$&$k_z$&10dB&40dB&10dB&40dB&10dB&40dB\\
\hline
$1$&$1$&$1$&\multicolumn{6}{c|}{\!1.000\ \ 1.000\ \ 0.994\ \ 1.000\ \ 0.991\ \ 0.997\!\!}\\
\cline{1-3}
$1,3$&$2$&$1,3$&\multicolumn{6}{c|}{\!1.000\ \ 1.000\ \ 0.996\ \ 0.997\ \ 0.989\ \ 0.991\!\!}\\
\cline{1-3}
$\!1,3,5,7$&$4$&$\!1,3,5,7$&\multicolumn{6}{c|}{\!1.000\ \ 1.000\ \ 0.997\ \ 0.997\ \ 0.986\ \ 0.988\!\!}\\
\cline{1-3}
\multicolumn{3}{|c|}{$\vdots$}&\multicolumn{6}{c|}{$\vdots$}\\
\cline{1-3}
$\!\!1,\!\cdots\!,\!127$&$64$&$\!\!1,\!\cdots\!,\!127$&\multicolumn{6}{c|}{\!0.996\ \ 0.996\ \ 1.000\ \ 1.000\ \ 1.000\ \ 1.000\!\!}\\
\cline{1-3}
$\!\!1,\!\cdots\!,\!255$&$128$&$\!\!1,\!\cdots\!,\!255$&\multicolumn{6}{c|}{\!0.993\ \ 0.993\ \ 1.000\ \ 1.000\ \ 1.000\ \ 1.000\!\!}\\
\hline
\end{tabular}}
\captionsetup{font={small}}
\caption{Identification accuracy for a UPA system with $N_x=N_z=64$ and $f=28$ GHz. The set $\mathcal{V}$ comprises grid points sampled on the plane at $y = -k_yD_x$. The value of $\hat{y}$ is set as $\hat{y}=(1-k_{\hat{y}})0.62\sqrt{\frac{D^3}{\lambda}}+k_{\hat{y}}\frac{2D^2}{\lambda}$, which lies between the Fresnel and Rayleigh distances. The reference SNR $\gamma$ is set to either 10 dB or 40 dB.}\label{tab1}
\vspace{-0.2cm}
\end{table}
\begin{figure*}[t]
    \centering
    \includegraphics[width=7in]{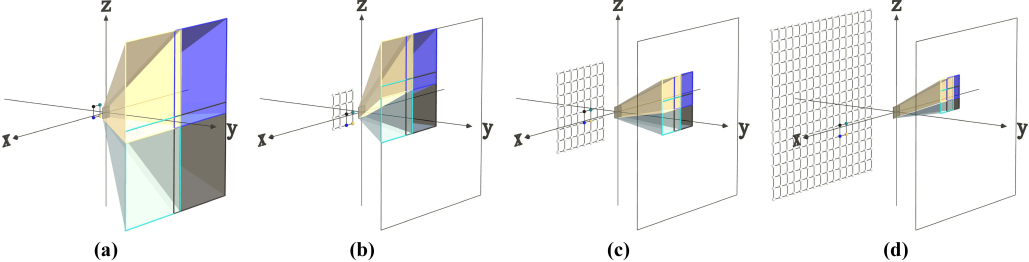}
    \captionsetup{font={small}}
    \caption{Illustration of the proposed hierarchical localization method. The spheres denote the virtual focal points in $\mathcal{V}_m$. (a) $\mathcal{R}(\bm{v}_{1,x,z})$ with $x,z\in\{1,2\}$; (b) $\mathcal{R}(\bm{v}_{2,x,z})$ with $x,z\in\{1,2\}$; (c) $\mathcal{R}(\bm{v}_{3,x,z})$ with $x,z\in\{3,4\}$; (d) $\mathcal{R}(\bm{v}_{4,x,z})$ with $x,z\in\{5,6\}$.}\label{fig:IV_method1_2}
    \vspace{-0.2cm}
\end{figure*}

The identification accuracy reflects the feasibility of leveraging diverging codewords to approximately localize the UE, which, if sufficiently high, can significantly aid beam training. Table \ref{tab1} presents the identification accuracy evaluated under various scenarios. The results show that it consistently approaches 1 across different settings of $\mathcal{V}$, $\hat{y}$, and $\gamma$, demonstrating the effectiveness of diverging codewords for UE localization and beam training. 

Two other key characteristics of the diverging codewords are the \textbf{controllable broad coverage} and \textbf{single RF chain requirement}. The former refers to their ability to flexibly control the region of high received pilot power, i.e., $\mathcal{R}_{\hat{y}}(\bm{v})$, by adjusting the position of the virtual focal point $\bm{v}$. The latter arises from the fact that diverging codewords apply only phase shifts across antennas without modifying their amplitudes, allowing implementation at the BS using a single RF chain. These characteristics, along with the previously validated high identification accuracy, make diverging codewords a promising solution for effective and efficient beam training.

\section{Beam Training with Diverging Codewords}

In this section, we utilize diverging codewords to design a two-phase beam training method in UPA systems. In the first phase, we design a hierarchical method to quickly locate the UE within a narrow, unbounded rectangular pyramidal frustum. In the second phase, a set of NF beam focusing codewords is deployed as pilots to achieve fine-grained alignment with the channel condition, thereby completing the beam training.

\subsection{Phase I: Hierarchical Localization}\label{sec:hie}

Motivated by the high identification accuracy of diverging codewords, we construct $M$ sets of virtual focal points satisfying the constraint in \eqref{eq:accuracy} to hierarchically narrow down the possible region containing the UE. Specifically, these point sets and their corresponding codebooks are defined as follows.
\begin{Def}[Diverging Codebook for UPA]
Given any $m\in\{1,2,\cdots\}$, we construct a set of virtual focal points as 
\begin{align*}
\mathcal{V}_m\triangleq\{\bm{v}_{m,x,z}|x,z\in\{1,2,\cdots,2^m\}\},
\end{align*}
where the point $\bm{v}_{m,x,z}$ is located at
\begin{align}
\bm{v}_{m,x,z}\!\!\triangleq\!\!\Big(\!\frac{2x\!\!-\!\!2^m\!\!\!-\!\!1}{2}\!D\!_x,-2^{m\!-\!1}\!\min\{D\!_x,\!D\!_z\},\!\frac{2z\!\!-\!\!2^m\!\!\!-\!\!1}{2}\!D\!_z\!\!\Big)\!.\label{def:vpoint}
\end{align}
Then, the (level-$m$) diverging codebook for UPA is defined as
\begin{align*}
\begin{split}
\bm{W}_m\triangleq&[\bm{c}(\bm{v}_{m,1,1}),\bm{c}(\bm{v}_{m,1,2}),\cdots,\bm{c}(\bm{v}_{m,1,2^m}),\\
&\qquad\qquad\qquad\quad\vdots\\
&\bm{c}(\bm{v}_{m,2^m,1}),\bm{c}(\bm{v}_{m,2^m,2}),\cdots,\bm{c}(\bm{v}_{m,2^m,2^m})].
\end{split}
\end{align*}
\end{Def}

It can be easily verified that the constraint in \eqref{eq:accuracy} is satisfied for every $\mathcal{V}_m$ with $m\in\{1,2,\cdots\}$ and every $\hat{y}>0$. Therefore, we can sequentially transmit all codewords in $\bm{W}_m$ as pilots and then identify which $\mathcal{R}(\bm{v}_{m,x,z})$ with $x,z\in\{1,2,\cdots,2^m\}$ contains the UE by comparing the powers of the received pilots. Here, $\mathcal{R}(\bm{v}_{m,x,z})$ denotes an unbounded rectangular pyramidal frustum defined by 
\begin{align}\label{def:frustum}
\mathcal{R}(\bm{v}_{m,x,z})\triangleq\{\bm{p}_{\bn{U}}\in\mathcal{R}_{\hat{y}}(\bm{v}_{m,x,z})|\hat{y}>0\}.
\end{align}
However, this method requires $4^m$ pilots, which becomes highly inefficient when $m$ is large---that is, when the objective is to localize the UE within a sufficiently narrow frustum.

To tackle this issue, we propose a hierarchical localization method. Specifically, we first transmit the four codewords in $\bm{W}_1$ as pilots and identify which $\mathcal{R}(\bm{v}_{1,x,z})$ contains the UE by comparing the powers of the four received pilots. Suppose the UE lies within $\mathcal{R}(\bm{v}_{1,x_1,z_1})$. Then, it can be theoretically validated that $\mathcal{R}(\bm{v}_{1,x_1,z_1})=\cup_{x\in\{2x_1-1,2x_1\},z\in\{2z_1-1,2z_1\}}\mathcal{R}(\bm{v}_{2,x,z})$ holds. Therefore, we can proceed by transmitting the four codewords $\{\bm{c}(\bm{v}_{2,x,z})|x\in\{2x_1-1,2x_1\},z\in\{2z_1-1,2z_1\}\}$ in $\bm{W}_2$ as pilots and locating the UE within one of the $\mathcal{R}(\bm{v}_{2,x,z})$ by comparing the received pilot powers again. By repeating this process for $M$ iterations, where $M$ is a given positive integer, the UE can eventually be localized within one of the $4^M$ narrow, unbounded rectangular pyramidal frustums $\mathcal{R}(\bm{v}_{M,x,z})$, where $x,z\in\{1,2,\cdots,2^M\}$, and the total pilot overhead is only $4M$. An illustrative example is provided in Fig. \ref{fig:IV_method1_2} for better clarity.
\begin{figure}[!t]
    \centering
    \includegraphics[width=0.87\columnwidth]{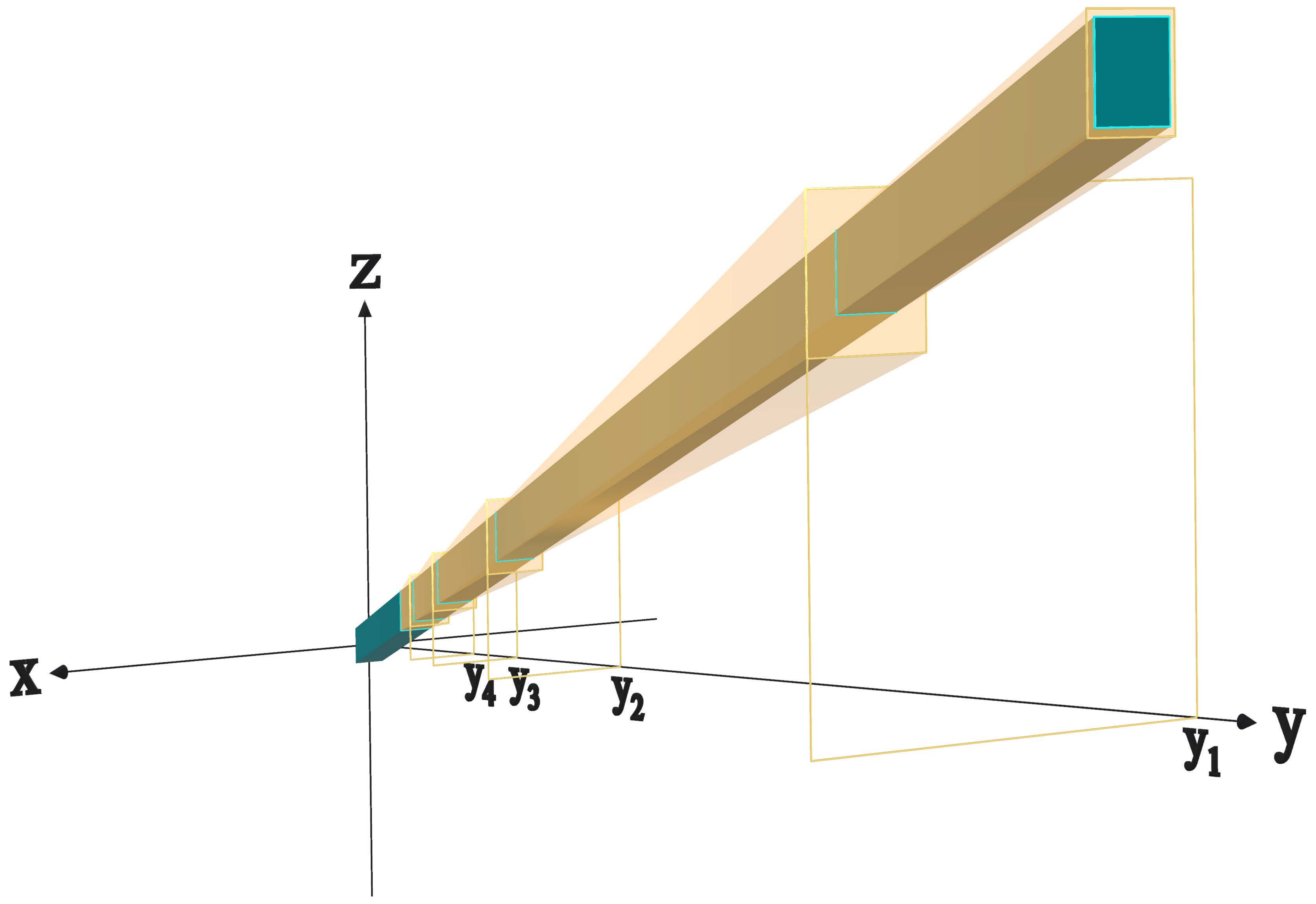}
    \captionsetup{font={small}}
    \caption{Illustration of the proposed NF refinement method. The grey block is a rectangular pyramidal frustum $\mathcal{R}(\bm{v}_{6,20,20})$. The orange blocks denote the expanded regions encompassing all potential focusing points to be sampled within $\mathcal{R}(\bm{v}_{6,20,20})$.}\label{fig:IV_method1_3}
    \vspace{-0.2cm}
\end{figure}

\subsection{Phase II: NF Refinement}\label{sec:refine}
After identifying the UE within a narrow frustum, we construct an NF refinement codebook, denoted by $\bm{W}_{\bn{R}}$, to match to the channel condition in a fine-grained manner. Specifically, this codebook consists solely of NF beam focusing codewords, each concentrating beam energy on a specific point. Based on the definition of the diverging codeword in \eqref{eq:mword}, the codeword focusing on point $\bm{u}$ can be represented as $\overline{\bm{c}(\bm{u})}$. Thus, the design of $\bm{W}_{\bn{R}}$ is equivalent to determining how to sample the focusing point $\bm{u}$ within the identified narrow frustum. Here, we use the triplet $(y_{\bm{u}},\!\phi_{\bm{u}},\!\theta_{\bm{u}})$ to represent the location of point $\bm{u}$, where $y_{\bm{u}}$, $\phi_{\bm{u}}$ and $\theta_{\bm{u}}$ are defined similarly to $y_{\bn{U}}$, $\phi_{\bn{U}}$ and $\theta_{\bn{U}}$ for point $\bm{p}_{\bn{U}}$.

As existing literature has not revealed well-structured orthogonality among NF beam focusing codewords targeting different points in UPA systems, we propose a heuristic method to sample $\bm{\bm{u}}$. Specifically, considering the high sensitivity of NF beam focusing codewords to angular misalignment, we adopt a dense sampling strategy over the angular dimensions $\phi_{\bm{u}}$ and $\theta_{\bm{u}}$, such that their tangent values lie in the set $\{\frac{2\times1-1}{2^M}-1,\frac{2\times2-1}{2^M}-1,\cdots\frac{2\times2^M-1}{2^M}-1\}$. Based on this strategy, the following lemma holds.

\begin{Lem}\label{lemma}
For any point $\bm{u}$ located within the frustum $\mathcal{R}(\bm{v}_{M,x_M,z_M})$, if $y_k< y_{\bm{u}}\leq y_{k-1}$ holds, then it follows that
\begin{align}
\begin{split}
\phi_{\bm{u}}\!\!\in&\Big\{\!\!\arctan\!\Big(\frac{2x\!-\!1\!}{2^M}-1\Big)\!\Big|\max\{1,\lfloor f_{\textnormal{min}}(D_x,x_M,k)\rfloor\}\!\!\leq\\
&\ \ x\leq \min\{2^M,\lceil f_{\textnormal{max}}(D_x,x_M,k)\rceil\}\Big\},\label{eq:phi}
\end{split}\\
\begin{split}
\theta_{\bm{u}}\!\!\in&\Big\{\!\!\arctan\!\Big(\frac{2z\!-\!1\!}{2^M}-1\Big)\!\Big|\max\{1,\lfloor f_{\textnormal{min}}(D_z,z_M,k)\rfloor\}\!\!\leq\\
&\ \ z\leq \min\{2^M,\lceil f_{\textnormal{max}}(D_z,z_M,k)\rceil\}\Big\},\label{eq:theta}
\end{split}
\end{align}
where
\begin{align}
\begin{split}
&\!f_{\textnormal{min}}(\!D\!,t,k\!)\!\!\triangleq\!\frac{\!(\min\{\!D_x,\!D_z\!\}\!+\!D\!)(2^M\!\!+\!\!1\!)}{2\min\{\!D_x,D_z\!\}}-\!\frac{D(t\!+\!k)}{\!\min\{\!D_x,\!D_z\!\}\!},
\end{split}\\
\begin{split}
&\!f_{\textnormal{max}}(\!D\!,t,k\!)\!\!\triangleq\!\frac{\!(\min\{\!D_x,\!D_z\!\}\!+\!D\!)(2^M\!\!+\!\!1\!)}{2\min\{\!D_x,D_z\!\}}-\!\frac{D(t\!-\!k)}{\!\min\{\!D_x,\!D_z\!\}\!},
\end{split}\\
&\!y_k\!\triangleq\left\{\begin{array}{ll}
\frac{2^{M\!-\!1}}{2k\!-\!1}\!\min\{\!D_x,\!D_z\!\}&k=1,2,\cdots\\
+\infty&k=0.
\end{array}\right.\label{def:yk}
\end{align}
\end{Lem}
The proof for Lemma \ref{lemma} is provided in Appendix \ref{app:II}. Based on this lemma, the angular sampling ranges of $\phi_{\bm{u}}$ and $\theta_{\bm{u}}$ vary with the value of $y_{\bm{u}}$. Consequently, for sampling purposes, the identified frustum $\mathcal{R}(\bm{v}_{M,x_M,z_M})$ can be equivalently expanded as the union of a series of rectangular pyramidal frustums, each bounded by different angular ranges over $\phi_{\bm{u}}$ and $\theta_{\bm{u}}$. An illustrative example is provided in Fig. \ref{fig:IV_method1_3} for better clarity. Moreover, noting that NF beam focusing codewords become increasingly sensitive to distance misalignment as the UE moves closer to the BS, the value of $y_{\bm{u}}$ is directly sampled from $\{y_k|k=1,2,\cdots\}$, since both the values of $y_k$ and the spacing between successive $y_k$ decrease as $k$ increases. Notably, intermittent sampling over $y_k$ (e.g., $k=1,3,5,\dots$ or $k=2,4,6,\dots$) can also be used, which reduces pilot overhead at the cost of some performance degradation.
\begin{figure}[t]
    \centering
    \includegraphics[width=0.8\columnwidth]{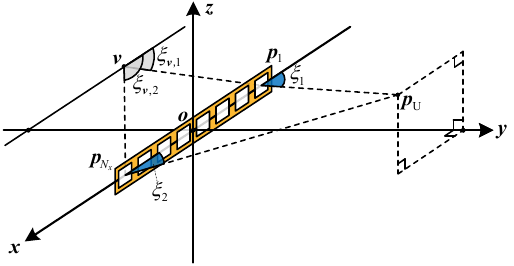}
    \captionsetup{font={small}}
    \caption{Horizontal ULA centered at the origin $\bm{o}$ and UE at $\bm{p}_{\bn{U}}$. \color{black}}\label{fig:V_system}
    \vspace{-0.2cm}
\end{figure}

The overall beam training method is summarized in Algorithm \ref{alg}, where the UE is localized within a narrow frustum in lines 2-6, followed by NF refinement in lines 7-8.

\begin{algorithm}[!b]
\caption{Two-phase NF beam training method for UPA}\label{alg}
\begin{algorithmic}[1]
\small
\STATE Denote the index of the identified frustum containing the UE among $\mathcal{R}(\bm{v}_{m,x,z})$ as $(x_m,z_m)$, and initialize $x_0$ and $z_0$ as 1;
\STATE \textbf{for} $m=1,2,\cdots,M$
\STATE \qquad Generate diverging codewords $\{\bm{c}(\!\bm{v}_{m,x,z}\!)|x\in\{2x_{m\!-\!1}\!-\!1,$
\STATEx \qquad $2x_{m\!-\!1}\},z\in\{2z_{m\!-\!1}\!-1,2z_{m\!-1}\}\}$ using \eqref{def:vpoint} and \eqref{eq:mword};
\STATE \qquad Transmit these codewords as pilots and derive the values of
\STATEx \qquad $x_m$ and $z_m$ by comparing the powers of the received pilots;
\STATE \qquad Report the values of $x_m$ and $z_m$ to the BS;
\STATE \textbf{end for};
\STATE Transmit all the NF focusing codewords with $\phi_{\bm{u}}$ and $\theta_{\bm{u}}$ sampled using \eqref{eq:phi} and \eqref{eq:theta}, and $y_{\bm{u}}$ sampled from $\{y_k|k=1,2,\cdots\}$;
\STATE Select the NF focusing codeword that yields the largest received pilot power.
\normalsize
\end{algorithmic}
\end{algorithm}

\section{An Alternative Beam Training Method}
Beyond the method proposed in the previous section, we are also interested in examining whether the beam diverging effect for ULA---thoroughly studied in our prior work \cite{li2025icc}---can be exploited for beam training in UPA systems and, if so, whether it can offer enhanced performance and at what additional cost. To this end, we revisit the beam diverging effect for ULA in this section and explore methods to leverage it for beam training.

\subsection{Beam Diverging Effect for ULA}
\begin{figure}[t]
    \centering
    \includegraphics[width=\columnwidth]{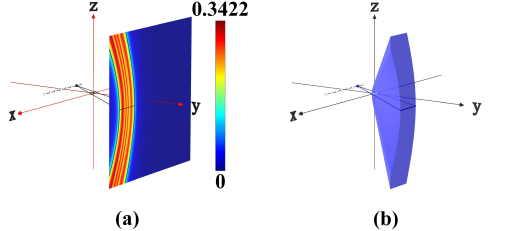}
    \captionsetup{font={small}}
    \caption{Beam diverging effect for ULA with $N_x=128$ and $\bm{v}=(-5D_x/2,-4D_x,0)$. (a) Amplitude of the received pilot on $\mathcal{R}_{3D}$ when deploying $\bm{c}(\bm{v})$ as the pilot. The black curves on the sampled UE plane represent the boundaries of $\mathcal{R}_{3D}^{\bn{hor}}(\bm{v})$; (b) $\mathcal{R}^{\bn{hor}}(\bm{v})$.}\label{fig:V_method2_1}
    \vspace{-0.2cm}
\end{figure}

Without loss of generality, we consider a horizontal $N_x$-antenna ULA aligned along the $x$-axis, which corresponds to a special case of a UPA with $N_z=1$. The first and $N_x$-th antennas of this ULA are positioned at $\bm{p}_{1,1}\triangleq(-\frac{D_x}{2},0,0)$ and $\bm{p}_{N_x,1}\triangleq(\frac{D_x}{2},0,0)$, respectively. 
% \color{magenta}
For clarity, an illustration is provided in Fig.~\ref{fig:V_system}. 
% \color{black}
In our previous study \cite{li2025icc}, the beam diverging effect for ULA was characterized in a 2D coordinate system. With a few additional notations, we generalize this effect to 3D space as follows.
\begin{Obs}[Beam Diverging Effect for ULA]
When deploying the diverging codeword $\bm{c}(\bm{v})$ as the pilot (identical to that defined in \eqref{eq:mword} with $N_z = 1$), the normalized power of the received pilot is substantially higher when the UE lies on the planar region 
\begin{align}\label{def:planar}
\mathcal{R}_{\hat{y}}^{\bn{hor}}(\bm{v})\triangleq\{\bm{p}_{\bn{U}}\in\mathcal{R}|y_{\bn{U}}=\hat{y},\xi_1>\xi_{\bm{v},1},\xi_2<\xi_{\bm{v},2}\},
\end{align}
compared to when the UE lies on its complement $\bar{\mathcal{R}}_{\hat{y}}^{\bn{hor}}(\bm{v})\triangleq\mathcal{R}_{\hat{y}}\setminus\mathcal{R}_{\hat{y}}^{\bn{hor}}(\bm{v})$. Here, $\xi_1, \xi_2, \xi_{\bm{v},1}$, and $\xi_{\bm{v},2}$ denote the angles between the negative $x$-axis and the vectors $\overrightarrow{\bm{p}_{1,1}\bm{p}_{\bn{U}}}$, $\overrightarrow{\bm{p}_{N_x,1}\bm{p}_{\bn{U}}}$, $\overrightarrow{\bm{vp}_{1,1}}$, and $\overrightarrow{\bm{vp}_{N_x,1}}$, respectively.
\end{Obs}
Based on the above observation, we can characterize the 3D beam patterns of diverging codewords in ULA systems. Specifically, based on \eqref{def:planar}, it can be proved that the boundaries of the high beam-power region $\mathcal{R}_{\hat{y}}^{\bn{hor}}(\bm{v})$ are hyperbolas, with an example shown in Fig. \ref{fig:V_method2_1}(a). Moreover, we define $\mathcal{R}^{\bn{hor}}(\bm{v})$ as the 3D high beam-power region formed by stacking $\mathcal{R}_{\hat{y}}^{\bn{hor}}(\bm{v})$ over $\hat{y}>0$, i.e.,
\begin{align}\label{def:shell}
\mathcal{R}^{\bn{hor}}(\bm{v})\triangleq\{\bm{p}_{\bn{U}}\in\mathcal{R}_{\hat{y}}^{\bn{hor}}(\bm{v})|\hat{y}>0\}.
\end{align}
It can be verified that $\mathcal{R}^{\bn{hor}}(\bm{v})$ constitutes a partial shell of a cone whose axis coincides with the $x$-axis, as illustrated in Fig. \ref{fig:V_method2_1}(b). 

We now examine whether the beam diverging effect in ULA systems can be utilized for beam training in 3D space. Following the approach in Section \ref{sec:iden}, we assess this possibility by evaluating the identification accuracy. Notably, the definition of identification accuracy in ULA systems is the same as that in UPA systems as given in Definition \ref{def:iden}, except that the condition in \eqref{eq:accuracy} is modified as
\begin{align}\label{eq:accuracy2}
\cup_{n=1}^N\mathcal{R}_{\hat{y}}^{\bn{hor}}(\bm{v}_n)=\mathcal{R}_{\hat{y}}.
\end{align}
The numerical results in Table \ref{tab2} show that the identification accuracy consistently approaches 1 across various settings of $\mathcal{V}$, $\hat{y}$, and $\gamma$, demonstrating that the UE can be precisely localized within a conical shell by sweeping a set of diverging codewords in ULA systems. However, even if the UE is localized within a very thin conical shell, this shell still spans a broad region in 3D space due to its geometric structure. As a result, the subsequent NF refinement required for beam training would still incur a considerable pilot overhead and is therefore inefficient.

\begin{table}[t]
\vspace{0.1cm}
\centering
\renewcommand{\arraystretch}{1}
{\footnotesize
\begin{tabular}{|>{\centering\arraybackslash}p{1.6cm}|>{\centering\arraybackslash}p{0.4cm}|>{\centering\arraybackslash}p{0.52cm}|>{\centering\arraybackslash}p{0.52cm}|>{\centering\arraybackslash}p{0.52cm}|>{\centering\arraybackslash}p{0.52cm}|>{\centering\arraybackslash}p{0.52cm}|>{\centering\arraybackslash}p{0.52cm}|}
\hline
\multicolumn{2}{|c|}{$\mathcal{V}\!\!=\!\!\{\!(\!\pm\!\frac{k\!_x}{2}\!\!D\!_x\!,\!-\!k\!_y\!D\!_x\!,\!0)\!\}$}&\multicolumn{2}{c|}{$k_{\hat{y}}$=0}&\multicolumn{2}{c|}{$k_{\hat{y}}$=0.3}&\multicolumn{2}{c|}{$k_{\hat{y}}$=0.7}\\
\hline
$k_x$&$k_y$&10dB&40dB&10dB&40dB&10dB&40dB\\
\hline
$1$&$1$&\multicolumn{6}{c|}{0.999\ \ \ 1.000\ \ \ 0.993\ \ \ 1.000\ \ \ 0.990\ \ \ 0.996}\\
\cline{1-2}
$1,3$&$2$&\multicolumn{6}{c|}{1.000\ \ \ 1.000\ \ \ 0.996\ \ \ 1.000\ \ \ 0.993\ \ \ 1.000}\\
\cline{1-2}
$1,3,5,7$&$4$&\multicolumn{6}{c|}{1.000\ \ \ 1.000\ \ \ 0.998\ \ \ 1.000\ \ \ 0.991\ \ \ 0.998}\\
\cline{1-2}
\multicolumn{2}{|c|}{$\vdots$}&\multicolumn{6}{c|}{$\vdots$}\\
\cline{1-2}
$1,\cdots,127$&$64$&\multicolumn{6}{c|}{0.999\ \ \ 0.999\ \ \ 1.000\ \ \ 1.000\ \ \ 0.996\ \ \ 1.000}\\
\cline{1-2}
$1,\cdots,255$&$128$&\multicolumn{6}{c|}{0.999\ \ \ 0.999\ \ \ 0.999\ \ \ 1.000\ \ \ 0.977\ \ \ 1.000}\\
\hline
\end{tabular}}
\captionsetup{font={small}}
\caption{Identification accuracy for a ULA system with $N_x=64$ and $f=28$ GHz. The set $\mathcal{V}$ comprises points sampled along the line where $y = -k_yD_x$ and $z=0$.}\label{tab2}
\vspace{-0.2cm}
\end{table}

\subsection{Beam Training through Beam Diverging Effect for ULA}
Now, we adopt the beam diverging effect for ULA to design a low-overhead beam training method in UPA systems. To achieve this, we note that additional RF chains can be employed in UPA systems to activate only a subset of antennas. In particular, activating a single row or column effectively configures the UPA as a ULA. Based on this observation, an intuitive beam training approach is to sequentially configure the UPA as ULAs at different positions or with different orientations, and then exploit the beam diverging effect across them to localize the UE within several conical shells. Subsequently, the NF refinement needs to be performed only on the intersection region of these shells, thereby reducing pilot overhead. Here, we formalize this approach as a three-phase beam training method. In the first two phases, the UPA is sequentially configured as two orthogonal ULAs to jointly localize the UE within a rod-like region. In the final phase, NF refinement is performed to complete the beam training.

\noindent\textbf{Phase I: Horizontal ULA-based Hierarchical Localization}

In the first phase, we localize the UE within a thin conical shell whose axis coincides with the $x$-axis. To achieve this, we activate only the $(1,\lceil\frac{N_z}{2}\rceil)$-th to the $(N_x,\lceil\frac{N_z}{2}\rceil)$-th antennas of the UPA, thereby forming a central horizontal ULA with $N_x$ antennas.\footnote{When $N_z$ is even, the central horizontal ULA does not strictly coincide with the $x$-axis. However, the positional offset between them is generally minor and can be verified to have a negligible impact on beam training performance.} We then deploy diverging codewords as pilots over this ULA to localize the UE. Specifically, these diverging codewords are selected from $M$ diverging codebooks, which are defined as follows.
\begin{Def}[Diverging Codebook for the Central Horizontal ULA]\label{def:mbookula}
Given any $m\in\{1,2,\cdots\}$, we construct a set of virtual focal points as 
\begin{align*}
\mathcal{V}_m^{\bn{hor}}\triangleq\{\bm{v}_{m,x}^{\bn{hor}}|x\in\{1,2,\cdots,2^m\}\},
\end{align*}
where the point $\bm{v}_{m,x}^{\bn{hor}}$ is located at
\begin{align}\label{def:vpoint2}
\bm{v}_{m,x}^{\bn{hor}}\triangleq\Big(\frac{2x\!-\!2^m\!-\!1}{2}D\!_x,-2^{m\!-\!1}\!D\!_x,0\Big).
\end{align}
Then, the (level-$m$) diverging codebook for the central horizontal ULA is defined as
\begin{align*}
\begin{split}
\bm{W}_m^{\bn{hor}}\triangleq&[\bm{c}^{\bn{hor}}(\bm{v}_{m,1}^{\bn{hor}}),\bm{c}^{\bn{hor}}(\bm{v}_{m,2}^{\bn{hor}}),\cdots,\bm{c}^{\bn{hor}}(\bm{v}_{m,2^m}^{\bn{hor}})],
\end{split}
\end{align*}
where $\bm{c}^{\bn{hor}}(\bm{v})\in\mathbb{C}^{N\!_x\!N\!_z\!\times1}$ denotes the diverging codeword modified according to the activation status of the $N_xN_z$ antennas and is defined as
\begin{align}\label{eq:mword2}
\begin{split}
\!\![\bm{c}^{\bn{hor}}\!(\bm{v})]_{(x\!-\!1)N_z+z}\!\!\triangleq\!\left\{\!\!\!\!\begin{array}{ll}
\frac{1}{\sqrt{N_x}}e\!^{-j\frac{2\pi}{\lambda}\!\big\|\overrightarrow{\bm{vp}_{x,z}}\!\big\|}&z=\lceil\!\frac{N_z}{2}\!\rceil\\
0&\textnormal{otherwise},
\end{array}
\right.
\end{split}
\end{align}
with $[\bm{c}^{\bn{hor}}(\bm{v})]_{k}$ denoting the $k$-th element of $\bm{c}^{\bn{hor}}(\bm{v})$.
\end{Def}
It can be verified that the constraint in \eqref{eq:accuracy2} holds for every $\mathcal{V}_m^{\bn{hor}}$ with $m\in\{1,2,\cdots\}$ and every $\hat{y}>0$. Thus, sweeping over any diverging codebook $\bm{W}_m^{\bn{hor}}$ with $m\in\{1,2,\cdots\}$ can precisely localize the UE within a conical shell $\mathcal{R}^{\bn{hor}}(\bm{v}_{m,x}^{\bn{hor}})$. 

\begin{figure}[t]
    \centering
    \includegraphics[width=\columnwidth]{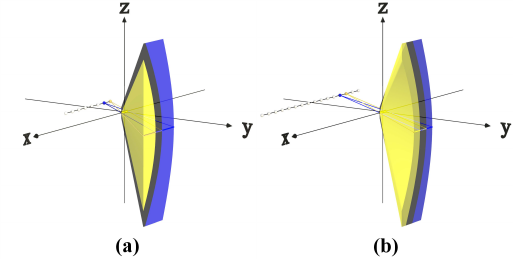}
    \captionsetup{font={small}}
    \caption{Illustrative examples of the proposed hierarchical localization method in Phase I. (a) Orange and blue spheres represent the points $\bm{v}_{3,1}^{\bn{hor}}$ and $\bm{v}_{3,2}^{\bn{hor}}$, respectively. Orange and blue regions represent $\mathcal{R}^{\bn{hor}}(\bm{v}_{3,1}^{\bn{hor}})$ and $\mathcal{R}^{\bn{hor}}(\bm{v}_{3,2}^{\bn{hor}})$ and the black region represents their intersection; (b) orange and blue spheres represent $\bm{v}_{4,3}^{\bn{hor}}$ and $\bm{v}_{4,4}^{\bn{hor}}$. Orange and blue regions represent $\mathcal{R}^{\bn{hor}}(\bm{v}_{4,3}^{\bn{hor}})$ and $\mathcal{R}^{\bn{hor}}(\bm{v}_{4,4}^{\bn{hor}})$.}\label{fig:V_method2_2}
    \vspace{-0.2cm}
\end{figure}

To localize the UE within the thinnest conical shell with minimal pilot overhead, we adopt a hierarchical method similar to that in Section \ref{sec:hie}. Specifically, we first transmit the two codewords in $\bm{W}_1^{\bn{hor}}$ as pilots and identify which $\mathcal{R}^{\bn{hor}}(\bm{v}_{1,x}^{\bn{hor}})$ contains the UE by comparing the two received pilot powers. Suppose $\mathcal{R}^{\bn{hor}}(\bm{v}_{1,x_1}^{\bn{hor}})$ contains the UE. We then transmit $\bm{c}^{\bn{hor}}(\bm{v}_{2,2x_1-1}^{\bn{hor}})$ and $\bm{c}^{\bn{hor}}(\bm{v}_{2,2x_1}^{\bn{hor}})$ as pilots and refine the localization to some $\mathcal{R}^{\bn{hor}}(\bm{v}_{2,x}^{\bn{hor}})$ by comparing the received powers again. After repeating this process for $M$ iterations, the UE is ultimately localized within some $\mathcal{R}^{\bn{hor}}(\bm{v}_{M,x}^{\bn{hor}})$, with a total pilot overhead of only $2M$. We provide illustrative examples of this hierarchical method in Fig. \ref{fig:V_method2_2}.

\noindent\textbf{Phase II: Vertical ULA-based Hierarchical Localization}

\begin{figure}[t]
    \centering
    \includegraphics[width=\columnwidth]{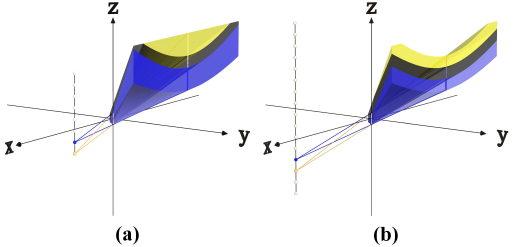}
    \captionsetup{font={small}}
    \caption{Illustrative examples of the proposed hierarchical localization method in Phase II. (a) Orange and blue spheres represent $\bm{v}_{3,1}^{\bn{ver}}$ and $\bm{v}_{3,2}^{\bn{ver}}$. Orange and blue regions represent $\mathcal{R}^{\bn{ver}}(\bm{v}_{3,1}^{\bn{ver}})$ and $\mathcal{R}^{\bn{ver}}(\bm{v}_{3,2}^{\bn{ver}})$; (b) orange and blue spheres represent $\bm{v}_{4,3}^{\bn{ver}}$ and $\bm{v}_{4,4}^{\bn{ver}}$. Orange and blue regions represent $\mathcal{R}^{\bn{ver}}(\bm{v}_{4,3}^{\bn{ver}})$ and $\mathcal{R}^{\bn{ver}}(\bm{v}_{4,4}^{\bn{ver}})$.}\label{fig:V_method2_3}
    \vspace{-0.2cm}
\end{figure}

In this phase, we localize the UE within a thin conical shell whose axis conincides with the $z$-axis. To achieve this, we activate only the $(\lceil\frac{N_x}{2}\rceil,1)$-th to the $(\lceil\frac{N_x}{2}\rceil,N_z)$-th antennas of the UPA, thereby forming a central vertical ULA with $N_z$ antennas. The diverging codewords deployed over this ULA are generated in parallel with those defined in Definition \ref{def:mbookula} and are thus omitted. The notations that serve as counterparts to $\mathcal{V}_m^{\bn{hor}}$, $\bm{v}_{m,x}^{\bn{hor}}$, $\bm{c}^{\bn{hor}}(\bm{v})$, $\bm{W}_m^{\bn{hor}}$, and $\mathcal{R}^{\bn{hor}}(\bm{v})$ are denoted as $\mathcal{V}_m^{\bn{ver}}$, $\bm{v}_{m,z}^{\bn{ver}}$, $\bm{c}^{\bn{ver}}(\bm{v})$, $\bm{W}_m^{\bn{ver}}$, and $\mathcal{R}^{\bn{ver}}(\bm{v})$, respectively, and their definitions are also omitted. The UE is also localized using a hierarchical method as in Phase I, with details omitted, and examples are illustrated in Fig. \ref{fig:V_method2_3}.

\noindent\textbf{Phase III: NF Refinement}
\begin{figure}[t]
    \centering
    \includegraphics[width=\columnwidth]{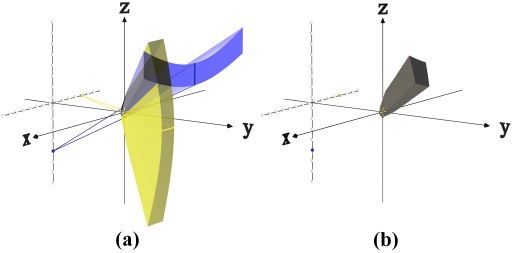}
    \captionsetup{font={small}}
    \caption{Illustrative example of the proposed hierarchical localization method with $M=4$. (a) Orange and blue regions represent $\mathcal{R}^{\bn{hor}}(\bm{v}_{4,4}^{\bn{hor}})$ and $\mathcal{R}^{\bn{ver}}(\bm{v}_{4,4}^{\bn{ver}})$, respectively; (b) the black region represents their intersection.}\label{fig:V_method2_4}
    \vspace{-0.2cm}
\end{figure}
\begin{figure}[!t]
    \centering
    \includegraphics[width=0.87\columnwidth]{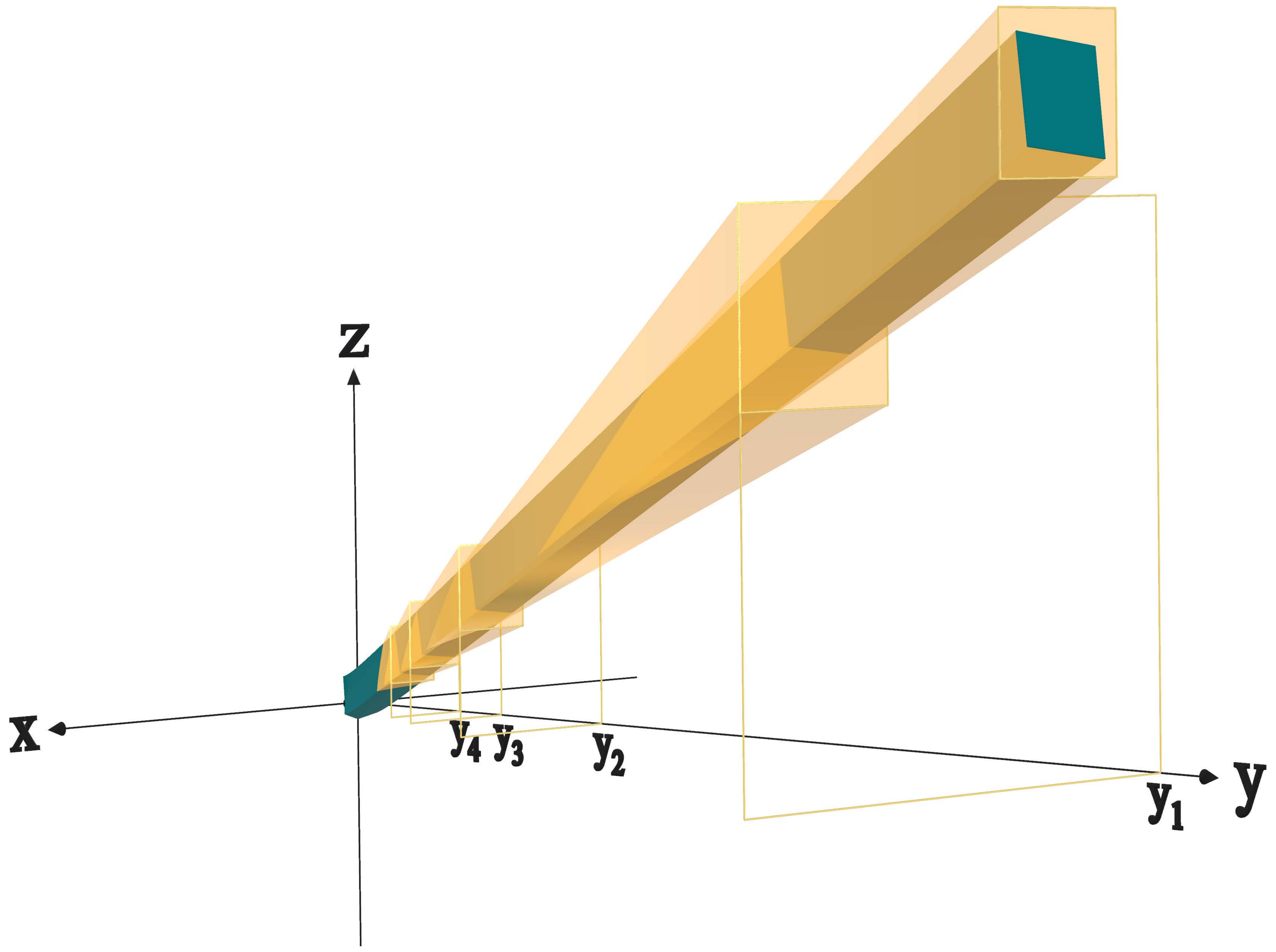}
    \captionsetup{font={small}}
    \caption{Illustrative example of the proposed NF refinement method in Phase III. The grey block is a rod-like region $\mathcal{R}^{\bn{hor}}(\bm{v}_{6,20}^{\bn{hor}})\cap\mathcal{R}^{\bn{ver}}(\bm{v}_{6,20}^{\bn{ver}})$. The orange blocks denote the expanded regions encompassing all potential focusing points to be sampled within $\mathcal{R}^{\bn{hor}}(\bm{v}_{6,20}^{\bn{hor}})\cap\mathcal{R}^{\bn{ver}}(\bm{v}_{6,20}^{\bn{ver}})$.}\label{fig:V_method2_5}
    \vspace{-0.2cm}
\end{figure}

Suppose the first two phases localize the UE within $\mathcal{R}^{\bn{hor}}(\bm{v}_{M,x_M}^{\bn{hor}})$ and $\mathcal{R}^{\bn{ver}}(\bm{v}_{M,z_M}^{\bn{ver}})$, respectively. Then, the UE must lie within their intersection $\mathcal{R}^{\bn{hor}}(\bm{v}_{M,x_M}^{\bn{hor}})\cap\mathcal{R}^{\bn{ver}}(\bm{v}_{M,z_M}^{\bn{ver}})$, which is a rod-like region as illustrated in Fig. \ref{fig:V_method2_4}.

Now, we operate NF refinement over this rod-like region to complete the beam training. This involves sampling a set of focusing points, each of which is represented by the triplet $\bm{u}=(y_{\bm{u}},\phi_{\bm{u}},\theta_{\bm{u}})$, as introduced in Section \ref{sec:refine}. We also adopt a heuristic sampling strategy over these three dimensions. Specifically, $y_{\bm{u}}$ is directly sampled from $\{y_k|k=1,2,\cdots\}$, as defined in \eqref{def:yk}, while $\phi_{\bm{u}}$ and $\theta_{\bm{u}}$ are again densely sampled such that their tangent values lie in the set $\{\frac{2\times1-1}{2^M}-1,\frac{2\times2-1}{2^M}-1,\cdots\frac{2\times2^M-1}{2^M}-1\}$. The sampling ranges of $\phi_{\bm{u}}$ and $\theta_{\bm{u}}$ should be as narrow as possible while fully covering the cross-sectional area of the rod-like region at $y=y_{\bm{u}}$. Apparently, these ranges vary with the value of $y_{\bm{u}}$ and can be easily calculated based on the indices $x_M$ and $z_M$. The calculation details are omitted for brevity. An illustration of this sampling method is provided in Fig. \ref{fig:V_method2_5} for better clarity.

We summarize this three-phase beam training method in Algorithm \ref{alg2}, where lines 2-11 correspond to the first two phases that localize the UE, and lines 12-13 correspond to the final NF refinement phase.

\begin{algorithm}[!t]
\caption{Three-phase NF beam training method for UPA}\label{alg2}
\begin{algorithmic}[1]
\small
\STATE Denote the index of the identified conical shell containing the UE among $\mathcal{R}^{\bn{hor}}(\bm{v}_{m,x}^{\bn{hor}})$ and $\mathcal{R}^{\bn{ver}}(\bm{v}_{m,z}^{\bn{ver}})$ as $x_m$ and $z_m$, respectively. Initialize $x_0$ and $z_0$ as 1;
\STATE \textbf{for} $m=1,2,\cdots,M$
\STATE \qquad Generate $\bm{c}^{\bn{hor}}\!(\!\bm{v}_{m\!,2x_{m\!-\!1}\!-\!1}^{\bn{hor}}\!)$ and $\bm{c}^{\bn{hor}}\!(\!\bm{v}_{m\!,2x_{m\!-\!1}}^{\bn{hor}}\!)$ using \eqref{def:vpoint2}\eqref{eq:mword2};
\STATE \qquad Transmit them as pilots and derive the values of $x_m$ by
\STATEx \qquad comparing the powers of the received pilots;
\STATE \qquad Report the values of $x_m$ to the BS;
\STATE \textbf{end for};
\STATE \textbf{for} $m=1,2,\cdots,M$
\STATE \qquad Generate $\bm{c}^{\bn{ver}}\!(\!\bm{v}_{m\!,2z_{m\!-\!1}\!-\!1}^{\bn{ver}}\!)$ and $\bm{c}^{\bn{ver}}\!(\!\bm{v}_{m\!,2z_{m\!-\!1}}^{\bn{ver}}\!)$;
\STATE \qquad Transmit them as pilots and derive the values of $z_m$ by
\STATEx \qquad comparing the powers of the received pilots;
\STATE \qquad Report the values of $z_m$ to the BS;
\STATE \textbf{end for};
\STATE Determine the sampled points based on the indices $x_M$, $z_M$. Transmit all corresponding NF focusing codewords as pilots;
\STATE Select the NF focusing codeword that yields the largest received pilot power.
\normalsize
\end{algorithmic}
\end{algorithm}

\begin{table*}[t]
\vspace{0.1cm}
\centering
\renewcommand{\arraystretch}{1}
{\footnotesize
\begin{tabular}{|
  >{\centering\arraybackslash}p{0.9cm}|
  p{0.1cm}|
  p{0.1cm}|
  p{0.1cm}|
  >{\centering\arraybackslash}p{0.8cm}|
  >{\centering\arraybackslash}p{0.8cm}|
  >{\centering\arraybackslash}p{0.8cm}|
  >{\centering\arraybackslash}p{0.8cm}|
  >{\centering\arraybackslash}p{0.8cm}|
  >{\centering\arraybackslash}p{0.8cm}|
  >{\centering\arraybackslash}p{0.8cm}|
  >{\centering\arraybackslash}p{0.8cm}|
  >{\centering\arraybackslash}p{0.8cm}|
  >{\centering\arraybackslash}p{0.8cm}|
  >{\centering\arraybackslash}p{0.8cm}|
  >{\centering\arraybackslash}p{0.8cm}|
}
\hline
\multirow{2}{*}{\hspace*{-0.6mm}\makecell[c]{Pilot\\overhead}}
& \multirow{2}{*}{\!\!\!$N_x$}
& \multirow{2}{*}{\!\!\!$N_z$}
& \multirow{2}{*}{\!\!\!$M$}
& \multicolumn{2}{c|}{\makecell[c]{Proposed two-\\phase method}} 
& \multicolumn{2}{c|}{\makecell[c]{Proposed three-\\phase method}} 
& \multicolumn{2}{c|}{\makecell[c]{UPA\\partitioning}} 
& \multicolumn{2}{c|}{\makecell[c]{ULA-based\\hierarchical DFT}} 
& \multicolumn{2}{c|}{\makecell[c]{ULA-based\\DFT sweeping}} 
& \multicolumn{2}{c|}{\makecell[c]{\hspace*{-1mm}Grid matching (la-\\\hspace*{-1mm}rge/small spacing)}}\\
\cline{5-16}
&&&&\hspace*{-1mm}{\scriptsize Phase I}&\hspace*{-1mm}{\scriptsize Phase II}&\hspace*{-1.6mm}{\scriptsize Phase I,II}&\hspace*{-1mm}{\scriptsize Phase III}&\hspace*{-1.6mm}{\scriptsize Phase I,II}&\hspace*{-1mm}{\scriptsize Phase III}&\hspace*{-1.6mm}{\scriptsize Phase I,II}&\hspace*{-1mm}{\scriptsize Phase III}&\hspace*{-1.6mm}{\scriptsize Phase I,II}&\hspace*{-1mm}{\scriptsize Phase III}&{\scriptsize Large}&{\scriptsize Small}\\
%\hline
%$1$&$1$&$1$&\multicolumn{6}{c|}{\!1.000\ \ 1.000\ \ 0.996\ \ 1.000\ \ 0.992\ \ 0.996\!\!}\\
\hline
\hspace*{-2mm}\multirow{2}{*}{\makecell[c]{Fig. \ref{fig:sim1},\\\ref{fig:sim2},\ref{fig:sim3}(b)}}
&64
&64
&9
&36&178&36&240&36&345&36&356&1024&359&\multirow{2}{*}{\textbf{125020}}&\multirow{2}{*}{\textbf{\!\!1000236}}\\
\cline{2-14}
&\multicolumn{3}{c|}{SUM}
&\multicolumn{2}{c|}{\textbf{214}}
&\multicolumn{2}{c|}{\textbf{276}}
&\multicolumn{2}{c|}{\textbf{381}}
&\multicolumn{2}{c|}{\textbf{392}}
&\multicolumn{2}{c|}{\textbf{1383}}
&&\\
\hline
\hspace*{-1mm}\multirow{2}{*}{Fig. \ref{fig:sim3}(a)}
&32
&32
&8
&32&57&32&86&32&223&32&235&512&237&\multirow{2}{*}{\textbf{15124}}&\multirow{2}{*}{\textbf{125020}}\\
\cline{2-14}
&\multicolumn{3}{c|}{SUM}
&\multicolumn{2}{c|}{\textbf{89}}
&\multicolumn{2}{c|}{\textbf{118}}
&\multicolumn{2}{c|}{\textbf{255}}
&\multicolumn{2}{c|}{\textbf{267}}
&\multicolumn{2}{c|}{\textbf{749}}
&&\\
\hline
\hspace*{-1mm}\multirow{2}{*}{Fig. \ref{fig:sim4}(a)}
&16
&32
&8
&32&45&32&79&32&200&32&235&512&237&\multirow{2}{*}{\textbf{4940}}&\multirow{2}{*}{\textbf{39546}}\\
\cline{2-14}
&\multicolumn{3}{c|}{SUM}
&\multicolumn{2}{c|}{\textbf{77}}
&\multicolumn{2}{c|}{\textbf{111}}
&\multicolumn{2}{c|}{\textbf{232}}
&\multicolumn{2}{c|}{\textbf{267}}
&\multicolumn{2}{c|}{\textbf{749}}
&& \\
\hline
\hspace*{-1mm}\multirow{2}{*}{Fig. \ref{fig:sim4}(b)}
&32
&64
&9
&36&105&36&160&36&204&36&238&1024&239&\multirow{2}{*}{\textbf{39546}}&\multirow{2}{*}{\textbf{301291}}\\
\cline{2-14}
&\multicolumn{3}{c|}{SUM}
&\multicolumn{2}{c|}{\textbf{141}}
&\multicolumn{2}{c|}{\textbf{196}}
&\multicolumn{2}{c|}{\textbf{240}}
&\multicolumn{2}{c|}{\textbf{274}}
&\multicolumn{2}{c|}{\textbf{1263}}
&& \\
\hline
\hspace*{-1mm}\multirow{2}{*}{Fig. \ref{fig:sim5}(a)}
&32
&32
&8
&32&99&32&141&32&222&32&237&512&239&\multirow{2}{*}{\textbf{45285}}&\multirow{2}{*}{\textbf{370916}}\\
\cline{2-14}
&\multicolumn{3}{c|}{SUM}
&\multicolumn{2}{c|}{\textbf{131}}
&\multicolumn{2}{c|}{\textbf{173}}
&\multicolumn{2}{c|}{\textbf{254}}
&\multicolumn{2}{c|}{\textbf{269}}
&\multicolumn{2}{c|}{\textbf{751}}
&& \\
\hline
\hspace*{-1mm}\multirow{2}{*}{Fig. \ref{fig:sim5}(b)}
&64
&64
&9
&36&260&36&359&36&474&36&499&1024&503&\multirow{2}{*}{\textbf{368999}}&\multirow{2}{*}{\textbf{\!\!2985203}}\\
\cline{2-14}
&\multicolumn{3}{c|}{SUM}
&\multicolumn{2}{c|}{\textbf{296}}
&\multicolumn{2}{c|}{\textbf{395}}
&\multicolumn{2}{c|}{\textbf{510}}
&\multicolumn{2}{c|}{\textbf{535}}
&\multicolumn{2}{c|}{\textbf{1527}}
&& \\
\hline
\hline
\multicolumn{4}{|c|}{Number of RF chains}
&\multicolumn{2}{c|}{1}
&\multicolumn{2}{c|}{3}
&\multicolumn{2}{c|}{$\log_2\!N_xN_z$}
&\multicolumn{2}{c|}{$\log_2\!N_xN_z$}
&\multicolumn{2}{c|}{3}
&\multicolumn{2}{c|}{1}\\
\hline
\end{tabular}}
\captionsetup{font={small}}
\caption{Simulated Pilot overhead and required number of RF chains for different beam training methods under various simulated scenarios.}\label{tab3}
\vspace{-0.2cm}
\end{table*}

\section{Numerical Results}\label{sec:num}
In this section, we numerically evaluate the beam training performance of the proposed methods. Specifically, we simulate a system with $N_x=N_z=64$ and $f=28$ GHz, yielding a Rayleigh distance of $42.525$ m. We assume a channel with one LOS path and eight NLOS paths, i.e., $L=8$, and set the Rician factor $|g_0|^2/\sum_{l=1}^L|g_l|^2$ to 13 dB. We set $M=9$ and define six distances, $\{y_k\}_{k=1}^6 = \{7.9, 9.6, 12.3, 17.3, 28.8, 86.4\}$ m, for NF refinement. To reduce pilot overhead, only $y_2$, $y_4$, and $y_6$ are sampled in the simulations, resulting in minimal performance degradation, as will be demonstrated shortly. For comparison, we consider the following benchmarks.
\begin{itemize}
\item \textbf{UPA Partitioning} \cite{partition}: This method comprises three phases. The first phase includes $M$ iterations. In the $m$-th iterations, the central $\min\{1,2^{\log_2N_z-m}\}$ horizontal ULAs within the UPA are activated, and two DFT-based codewords are sequentially deployed over them to halve the azimuth search range. The second phase adopts a similar procedure to estimate the UE's elevation. The final phase performs NF refinement within a spatial region centered on the direction determined by the estimated azimuth and elevation, and samples $y_{\bm{u}}$ from $\{y_k\}_{k=1}^6$.
\item \textbf{ULA-based hierarchical DFT} \cite{YouBeamTraining}: This method also comprises three phases. In the first two phases, hierarchical DFT codewords are applied to the central horizontal and vertical ULAs within the UPA to estimate the azimuth and elevation, respectively. The final phase also performs NF refinement.
\item \textbf{ULA-based DFT sweeping}: This method estimates the azimuth and elevation by sweeping DFT codewords across the central horizontal and vertical ULAs, and subsequently performs NF refinement.
\item \textbf{Grid Matching} \cite{NFris}: This method applies NF focusing beams to a 3D spatial grid with uniform spacing along the $x$-, $y$-, and $z$-axes. 
\end{itemize}
We evaluate these algorithms by examining their SNR loss, defined as the difference between the theoretical SNR upper bound under full array gain, $\frac{N_x N_z \sum_{l=0}^L |g_l|^2}{\sigma^2}$, and the SNR actually achieved. The quantity $\frac{\sum_{l=0}^L |g_l|^2}{\sigma^2}$ is referred to as the reference SNR, which is varied to simulate different channel gains or noise power levels.

The pilot overhead and the number of required RF chains for various algorithms are summarized in Table \ref{tab3}. The two proposed methods require lower pilot overhead compared to the benchmark methods. In addition, they require very few RF chains: the two-phase method uses only one, while the three-phase method uses three.
\begin{figure}[t]
    \centering
    \includegraphics{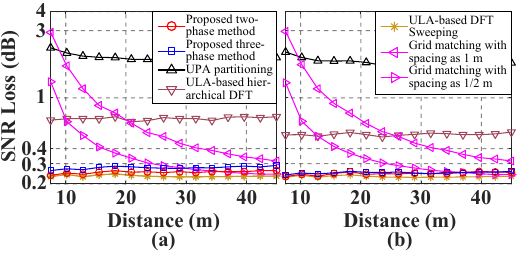}
    \captionsetup{font={small}}
    \caption{Distance v.s. SNR loss of different beam training methods at reference SNR levels of: (a) 15 dB; (b) 35 dB.}\label{fig:sim1}
    \vspace{-0.2cm}
\end{figure}

\begin{figure}[t]
    \centering
    \includegraphics{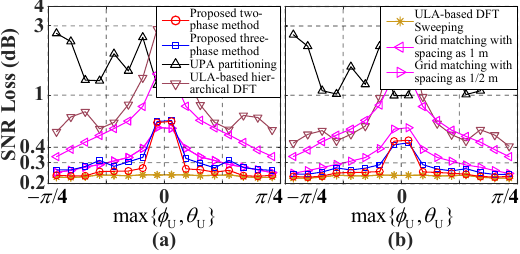}
    \captionsetup{font={small}}
    \caption{$\max\{\phi_{\bn{U}},\theta_{\bn{U}}\}$ v.s. SNR lossS of different beam training methods at reference SNR levels of: (a) 15 dB; (b) 35 dB.}\label{fig:sim2}
    \vspace{-0.2cm}
\end{figure}

\begin{figure}[t]
    \centering
    \includegraphics{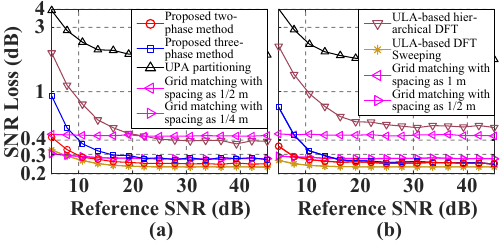}
    \captionsetup{font={small}}
    \caption{Reference SNR v.s. SNR loss of different beam training methods in scenarios with: (a) $N_x=N_z=32$ (Rayleigh distance: $10.296$\textnormal{ m}), and UEs uniformly sampled from $\{\bm{p}_{\bn{U}}\in\mathcal{R}_{\hat{y}}|\hat{y}\in[1.875,11.25)\textnormal{ m}\}$; (b) $N_x=N_z=64$, and UEs uniformly sampled from $\{\bm{p}_{\bn{U}}\in\mathcal{R}_{\hat{y}}|\hat{y}\in[7.5,45)\textnormal{ m}\}$.}\label{fig:sim3}
    \vspace{-0.4cm}
\end{figure}

\begin{figure}[t]
    \centering
    \includegraphics{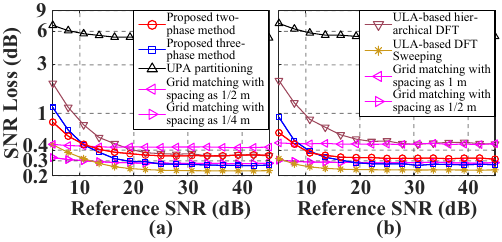}
    \captionsetup{font={small}}
    \caption{Reference SNR v.s. SNR loss of different beam training methods in scenarios with: (a) $N_x=16$, $N_z=32$ (Rayleigh distance: $6.354$\textnormal{ m}), and UEs uniformly sampled from $\{\bm{p}_{\bn{U}}\in\mathcal{R}_{\hat{y}}|\hat{y}\in[1.25,7.5)\textnormal{ m}\}$; (b) $N_x=32$, $N_z=64$ (Rayleigh distance: $26.411$\textnormal{ m}), and UEs uniformly sampled from $\{\bm{p}_{\bn{U}}\in\mathcal{R}_{\hat{y}}|\hat{y}\in[5,30)\textnormal{ m}\}$.}\label{fig:sim4}
    \vspace{-0.2cm}
\end{figure}

\begin{figure}[t]
    \centering
    \includegraphics{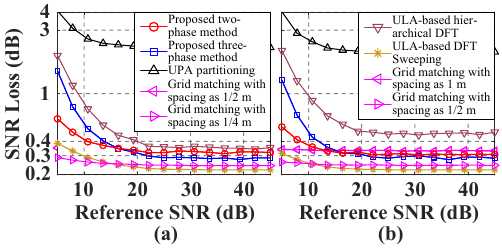}
    \captionsetup{font={small}}
    \caption{Reference SNR v.s. SNR loss of different beam training methods in scenarios where UEs are sampled from $\{\bm{p}_{\bn{U}}|y_{\bn{U}}\!>\!0,-\frac{\pi}{3}\!<\!\phi_{\bn{U}}\!<\!\frac{\pi}{3},-\frac{\pi}{3}\!<\!\theta_{\bn{U}}\!<\!\frac{\pi}{3}\}$, with: (a) $N_x=N_z=32$; (b) $N_x=N_z=64$.}\label{fig:sim5}
    \vspace{-0.2cm}
\end{figure}

Fig. \ref{fig:sim1} presents simulations where UEs are sampled from planes at varying distances from the UPA, i.e., from regions $\mathcal{R}_{\hat{y}}$ with different $\hat{y}$. 
The two proposed methods achieve a stable SNR loss across varying distances. Their SNR loss is also very close to that of the ULA-based DFT sweeping method, approaching the theoretical lower bound of 0.22 dB at a Rician factor of 13 dB, and substantially lower than those of the other benchmarks.
In contrast, the UPA partitioning method suffers from a large SNR loss because the ULAs within the partitioned UPA simultaneously transmit different DFT beams, each with a sigmoid-shaped profile in the angular domain. As a result, the side lobes of these beams can significantly reinforce one another, distorting angular estimation and degrading the overall SNR performance. 
The ULA-based hierarchical DFT method also relies on DFT beams and performs well only under high-reference SNR conditions. This is because, in low-reference SNR regimes, the side lobes of these DFT beams become more pronounced, again leading to distorted angular estimation and degraded SNR. 
The grid matching method underperforms at close ranges, where coarse grid resolution leads to poor angular estimation performance. 
In summary, the two proposed methods deliver strong and robust SNR performance across varying distances. While the ULA-based DFT sweeping method achieves a slightly smaller SNR loss, it does so at the cost of substantially higher pilot overhead.

In Fig. \ref{fig:sim2}, we simulate a scenario where UEs are sampled over different values of angles $\phi_{\bn{U}}$ and $\theta_{\bn{U}}$, and plot $\max\{\phi_{\bn{U}}, \theta_{\bn{U}}\}$ versus SNR loss for various methods. 
It is observed that most algorithms exhibit a relatively large SNR loss when $\max\{\phi_{\bn{U}}, \theta_{\bn{U}}\}$ is near 0, indicating that the angular misalignment sensitivity is strongest around the direction pointing toward $+y$. 
The two proposed algorithms again achieve relatively low and stable SNR loss across the entire angular domain, demonstrating their robustness to variations in UE direction.

In Fig. \ref{fig:sim3}, we simulate scenarios with different UPA sizes. The two proposed methods consistently achieve low SNR loss, demonstrating robustness to UPA size variations. 
In Fig. \ref{fig:sim4}, we further consider non-square UPA shapes. In such cases, the virtual focal points generated by the proposed two-phase method can become too close to the UPA, leading to reduced identification accuracy (validated to fall below 0.9) and a slight degradation in SNR. In contrast, the proposed three-phase method maintains identification accuracy close to 1 and continues to achieve low SNR loss across different reference SNR levels.
Fig. \ref{fig:sim5} examines scenarios where the UE region is expanded to $\{\bm{p}_{\bn{U}}|y_{\bn{U}}\!>\!0,-\frac{\pi}{3}\!<\!\phi_{\bn{U}}\!<\!\frac{\pi}{3},-\frac{\pi}{3}\!<\!\theta_{\bn{U}}\!<\!\frac{\pi}{3}\}$. In this setting, the proposed two-phase method again exhibits reduced identification accuracy (also fall below 0.9), resulting in a slight degradation in SNR. In contrast, the proposed three-phase method continues to maintain high identification accuracy and low SNR loss across all reference SNR levels. 

In conclusion, the proposed two methods demonstrate strong and robust SNR performance across varying UE distances, UE directions, SNR levels, UPA configurations, and UE regions. 
In particular, the two-phase method requires only a single RF chain and the lowest pilot overhead, remaining efficient even under very low SNR conditions. 
The three-phase method consistently achieves low SNR loss across most scenarios, including those with non-square UPAs or expanded UE regions.

\section{Conclusions and Future Work}
In this paper, we first investigated the beam diverging effect for UPAs by exploring methods to induce it and proposing a metric to evaluate it. We then leveraged this effect to design a two-phase beam training method, where the first phase localizes the UE within a narrow frustum with low pilot overhead, and the second phase performs fine-grained channel matching. Subsequently, we extended the beam diverging effect for ULA to 3D space and proposed a three-phase beam training method, which also achieves low pilot overhead and high-precision channel matching. Numerical results demonstrate the strong and robust performance of the two proposed methods. For future work, we plan to rigorously explore the theoretical foundations underlying the beam diverging effect and apply it to a broader range of NF tasks.
\appendices
\renewcommand{\thesection}{\Alph{section}} % Section A, B, ...
\renewcommand{\theThm}{\thesection.\arabic{Thm}}
\section{NF Region Characterization in UPA Systems}\label{app:I}
In this section, we first derive the mathematical expression for the UPA's NF region. Then, we simplify this expression and prove that the UPA's NF region is equivalent to the union of the NF regions of two ULAs. Finally, we derive the smallest hemisphere that serves as a superset of the UPA's NF region.

To derive the expression for the UPA's NF region, we first obtain the unit direction vector of the UE as $\hat{\bm{p}}_{\bn{U}}\triangleq\frac{\overrightarrow{\bm{op}_{\bn{U}}}}{\big\|\overrightarrow{\bm{op}_{\bn{U}}}\big\|}$. Using this vector, we then apply FF approximation to estimate the distance between the $(x,z)$-th UPA antenna and the UE as $\|\overrightarrow{\bm{p}_{x,z}\bm{p}_{\bn{U}}}\|\approx\overrightarrow{\bm{p}_{x,z}\bm{p}_{\bn{U}}}\cdot\hat{\bm{p}}_{\bn{U}}$, which induces a non-negative phase error of 
\begin{align}\label{def:beta}
\beta_{x,z}\triangleq\frac{2\pi}{\lambda}\left(\|\overrightarrow{\bm{p}_{x,z}\bm{p}_{\bn{U}}}\|-\overrightarrow{\bm{p}_{x,z}\bm{p}_{\bn{U}}}\cdot\hat{\bm{p}}_{\bn{U}}\right).	
\end{align}
According to the definition of the NF region, the UE lies within the UPA's NF region if and only if the maximum phase error among the $N_xN_z$ antenna-UE links is no less than $\frac{\pi}{8}$. Therefore, the UPA's NF region can be expressed as follows.
\begin{align*}
\mathcal{R}_{\bn{UPA}}\triangleq\Big\{\bm{p}_{\bn{U}}\Big|\max_{x\in\mathcal{N}_x,z\in\mathcal{N}_z}\beta_{x,z}\geq\frac{\pi}{8}\Big\}.
\end{align*}

Given the above expression, however, it remains difficult to understand the structure of the UPA's NF region, since it involves a maximum operation over $N_x N_z$ values. To tackle this issue, we introduce the following lemma.
\begin{Lem}\label{lemma1}
When the phase error $\beta_{x,z}$ defined in \eqref{def:beta} reaches its maximum over all $x\in\mathcal{N}_x$ and $z\in\mathcal{N}_z$, the conditions $x\in\{1,N_x\}$ and $z\in\{1,N_z\}$ hold. In other words, it follows that
\begin{align*}
\mathcal{R}_{\bn{UPA}}=\Big\{\bm{p}_{\bn{U}}\Big|\max_{x\in\{1,N_x\},z\in\{1,N_z\}}\beta_{x,z}\geq\frac{\pi}{8}\Big\}.
\end{align*}
\end{Lem}

\begin{IEEEproof}
To prove Lemma \ref{lemma1}, it is sufficient to show that for any antenna lying on the line segment connecting two other antennas, its corresponding phase error does not exceed the maximum of the phase errors at those two antennas. In other words, if $\bm{p}_{x_3,z_3}\triangleq\alpha\bm{p}_{x_1,z_1}+(1-\alpha)\bm{p}_{x_2,z_2}$ with $\alpha\in[0,1]$ holds, then the phase error $\beta_{x_3,z_3}$ satisfies
\begin{align}\label{eq:upa2}
\beta_{x_3,z_3}\leq\max\{\beta_{x_1,z_1},\beta_{x_2,z_2}\}.
\end{align}

\begin{figure}[tb]
\centering
\includegraphics[width=0.45\columnwidth]{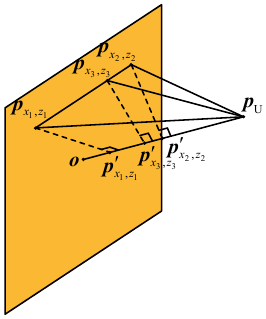}
\caption{Illustration demonstrating the proof of \eqref{eq:upa2}.}\label{fig:app_1}
\vspace{-0.2cm}
\end{figure}

To prove \eqref{eq:upa2}, we provide an illustration in Fig. \ref{fig:app_1}, where the points $\bm{p}'_{x_1,z_1}$, $\bm{p}'_{x_2,z_2}$, and $\bm{p}'_{x_3,z_3}$ represent the orthogonal projections of $\bm{p}_{x_1,z_1}$, $\bm{p}_{x_2,z_2}$, and $\bm{p}_{x_3,z_3}$ onto the line $\bm{op}_{\bn{U}}$, respectively. Based on the definition in \eqref{def:beta}, it follows that
\begin{align}
\beta_{x_i,z_i}&=\frac{2\pi}{\lambda}\left(\|\overrightarrow{\bm{p}_{x_i,z_i}\bm{p}_{\bn{U}}}\|-\overrightarrow{\bm{p}_{x_i,z_i}\bm{p}_{\bn{U}}}\cdot\hat{\bm{p}}_{\bn{U}}\right),\forall\ i\in\{1,2,3\},\nonumber
\end{align}
where the magnitude of
$\overrightarrow{\bm{p}_{x_i,z_i}\bm{p}_{\bn{U}}}\cdot\hat{\bm{p}}_{\bn{U}}$ equals $\|\overrightarrow{\bm{p}'_{x_i,z_i}\bm{p}_{\bn{U}}}\|$ for all $i\in\{1,2,3\}$. Note that $\overrightarrow{\bm{p}_{x_3,z_3}\bm{p}_{\bn{U}}}=\alpha\overrightarrow{\bm{p}_{x_1,z_1}\bm{p}_{\bn{U}}}+(1-\alpha)\overrightarrow{\bm{p}_{x_2,z_2}\bm{p}_{\bn{U}}}$ holds. Thus, we have
\begin{align*}
\beta_{x_3,z_3}=&\frac{2\pi}{\lambda}\Big(\|\alpha\overrightarrow{\bm{p}_{x_1,z_1}\bm{p}_{\bn{U}}}+(1-\alpha)\overrightarrow{\bm{p}_{x_2,z_2}\bm{p}_{\bn{U}}}\|\\
&-(\alpha\overrightarrow{\bm{p}_{x_1,z_1}\bm{p}_{\bn{U}}}+(1-\alpha)\overrightarrow{\bm{p}_{x_2,z_2}\bm{p}_{\bn{U}}})\cdot\hat{\bm{p}}_{\bn{U}}\Big)\nonumber\\
\leq&\frac{2\pi}{\lambda}\Big(\alpha\|\overrightarrow{\bm{p}_{x_1,z_1}\bm{p}_{\bn{U}}}\|+(1-\alpha)\|\overrightarrow{\bm{p}_{x_2,z_2}\bm{p}_{\bn{U}}}\|\\
&-(\alpha\overrightarrow{\bm{p}_{x_1,z_1}\bm{p}_{\bn{U}}}+(1-\alpha)\overrightarrow{\bm{p}_{x_2,z_2}\bm{p}_{\bn{U}}})\cdot\hat{\bm{p}}_{\bn{U}}\Big)\nonumber\\
%=&\frac{2\pi}{\lambda}\Big(\alpha\left(\|\overrightarrow{\bm{p}_{x_1,z_1}\bm{p}_{\bn{U}}}\|-\overrightarrow{\bm{p}_{x_1,z_1}\bm{p}_{\bn{U}}}\cdot\hat{\bm{p}}_{\bn{U}}\right)\\
%&+(1-\alpha)\left(\|\overrightarrow{\bm{p}_{x_2,z_2}\bm{p}_{\bn{U}}}\|-\overrightarrow{\bm{p}_{x_2,z_2}\bm{p}_{\bn{U}}}\cdot\hat{\bm{p}}_{\bn{U}}\right)\Big)\\
=&\alpha\beta_{x_1,z_1}+(1-\alpha)\beta_{x_2,z_2}\leq\max\{\beta_{x_1,z_1},\beta_{x_2,z_2}\},
\end{align*}
which completes the proof.
\end{IEEEproof}

Using Lemma \ref{lemma1}, it is easy to derive that
\begin{align}\label{eq:upa7}
\mathcal{R}_{\bn{UPA}}=\mathcal{R}_{(1,1),(N_x,N_z)}\cup\mathcal{R}_{(N_x,1),(1,N_z)},
\end{align}
where
\begin{align*}
\mathcal{R}_{(1,1),(N_x,N_z)}&\triangleq\Big\{\bm{p}_{\bn{U}}\Big|\max_{(x,z)\in\{(1,1),(N_x,N_z)\}}\beta_{x,z}\geq\frac{\pi}{8}\Big\},\\
\mathcal{R}_{(N_x,1),(1,N_z)}&\triangleq\Big\{\bm{p}_{\bn{U}}\Big|\max_{(x,z)\in\{(N_x,1),(1,N_z)\}}\beta_{x,z}\geq\frac{\pi}{8}\Big\}.
\end{align*}
It can be verified that $\mathcal{R}_{(1,1),(N_x,N_z)}$ and $\mathcal{R}_{(N_x,1),(1,N_z)}$ correspond to the NF regions of two ULAs with endpoints at $\bm{p}_{1,1}$ and $\bm{p}_{N_x,N_z}$, and at $\bm{p}_{N_x,1}$ and $\bm{p}_{1,N_z}$, respectively. Thus, equation \eqref{eq:upa7} indicates that the UPA's NF region is equivalent to the union of the NF regions of these two ULAs. 

Moreover, it can be proved that the boundary of $\mathcal{R}_{(1,1),(N_x,N_z)}$ can be approximately expressed as
\begin{align*}
\Big\{\bm{p}_{\bn{U}}\Big|r_{\bn{U}}=\frac{2D^2}{\lambda}\left(1-\frac{(D_x\cos\theta_{\bn{U}}\cos\phi_{\bn{U}}+D_z\sin\theta_{\bn{U}})^2}{D^2}\right)\Big\},
\end{align*}
and the boundary of $\mathcal{R}_{(N_x,1),(1,N_z)}$ as
\begin{align*}
\Big\{\bm{p}_{\bn{U}}\Big|r_{\bn{U}}=\frac{2D^2}{\lambda}\left(1-\frac{(D_x\cos\theta_{\bn{U}}\cos\phi_{\bn{U}}-D_z\sin\theta_{\bn{U}})^2}{D^2}\right)\Big\}.
\end{align*}
Therefore, the UPA's NF region is contained within the hemisphere $\mathcal{R}_{\bn{hem}}\!=\!\{\bm{p}_{\bn{U}}|y_{\bn{U}}\!>\!0,r_{\bn{U}}\!\leq\!\frac{2D^2}{\lambda}\}$. Furthermore, numerical validation shows that when $N_x=N_z$ holds, the NF region for UPA occupies more than 54\% of this hemisphere's volume.

\section{Proof of Lemma \ref{lemma}}\label{app:II}
\begin{figure}
\centering
\includegraphics[width=0.8\columnwidth]{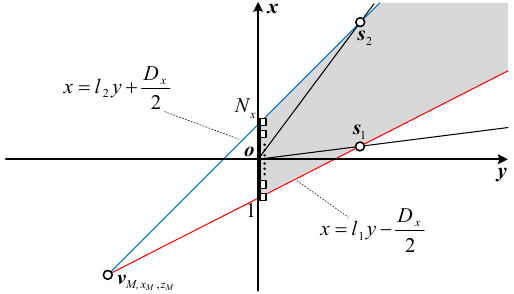}
\caption{Illustration of the proof of Lemma \ref{lemma}.}\label{fig:D_1}
\end{figure}

To prove Lemma \ref{lemma}, we begin by geometrically characterize the frustum $\mathcal{R}(\bm{v}_{M,x_M,z_M})$. Specifically, based on \eqref{def:rect}, \eqref{def:vpoint}, and \eqref{def:frustum}, if $\bm{p}_{\bn{U}}\in\mathcal{R}(\bm{v}_{M,x_M,z_M})$, then
\begin{align}\label{con}
l_1y_{\bn{U}}-\frac{D_x}{2}<x_{\bn{U}}<l_2y_{\bn{U}}+\frac{D_x}{2},
\end{align}
where 
\begin{align*}
l_1\triangleq \Big(\frac{2x_M\!\!-\!\!2^M\!\!-\!\!1}{2}D_x\!+\!\frac{D_x}{2}\!\Big)\!\frac{1}{-2^{M\!-\!1}\!\min\{\!D_x,\!D_z\!\}},\\
l_2\triangleq \Big(\frac{2x_M\!\!-\!\!2^M\!\!-\!\!1}{2}D_x\!-\!\frac{D_x}{2}\!\Big)\!\frac{1}{-2^{M\!-\!1}\!\min\{\!D_x,\!D_z\!\}}.
\end{align*}
Therefore, from a bottom-up view along the $z$-axis, the frustum $\mathcal{R}(\bm{v}_{M,x_M,z_M})$ is projected onto a 2D region bounded by the lines $x=l_1y-\frac{D_x}{2}$ and $x=l_2y+\frac{D_x}{2}$, along with the $x$-axis, as illustrated in Fig. \ref{fig:D_1}. 

Now, we prove \eqref{eq:phi} in Lemma \ref{lemma}. Denote the intersection point between the line $x=l_1y-\frac{D_x}{2}$ and the ray at angle of $\arctan(\frac{2f_{\textnormal{min}}(D_x,x_M,k)-1}{2^M}-1)$ as $\bm{s}_1\triangleq(y_{\bm{s}_1},x_{\bm{s}_1})$, where $k\in\{1,2,\cdots\}$. Then, we have
\begin{align}
x_{\bm{s}_1}\!\!&=\!l_1y_{\bm{s}_1}-\frac{D_x}{2},\label{eq:line}\\
x_{\bm{s}_1}\!\!&=y_{\bm{s}_1}\Big(\frac{2f_{\textnormal{min}}(D_x,x_M,k)-1}{2^M}-1\Big).\label{eq:ray}
\end{align}
By jointly solving \eqref{eq:line} and \eqref{eq:ray}, we obtain $y_{\bm{s}_1}=y_k$ for $k\in\{1,2,\cdots\}$. Similarly, denote the intersection point between $x=l_2y+\frac{D_x}{2}$ and the ray at angle of $\arctan(\frac{2f_{\textnormal{max}}(D_x,x_M,k)-1}{2^M}-1)$ as $\bm{s}_2\triangleq(y_{\bm{s}_2},x_{\bm{s}_2})$. We can also prove that $y_{\bm{s}_2}=y_k$. Therefore, \eqref{eq:phi} holds.

Using a similar approach, we can prove that \eqref{eq:theta} also holds, thereby completing the proof of Lemma \eqref{lemma}.

\bibliographystyle{IEEEtran}
\bibliography{refs}

\end{document}

%% file: acronyms.tex
\DeclareAcronym{ELAA}{
  short=ELAA,
  long=extremely large-scale antenna array
}

\DeclareAcronym{BS}{
  short=BS,
  long=base station
}

\DeclareAcronym{UE}{
  short=UE,
  long=user equipment
}

\DeclareAcronym{NF}{
  short=NF,
  long=near-field
}

\DeclareAcronym{FF}{
  short=FF,
  long=far-field
}

\DeclareAcronym{MPC}{
  short=MPC,
  long=mirror polar-domain codebook  
}

\DeclareAcronym{ULA}{
  short=ULA,
  long=uniform linear array
}

\DeclareAcronym{UPA}{
  short=UPA,
  long=uniform planar array
}

\DeclareAcronym{RF}{
  short=RF,
  long=radio frequency
}

\DeclareAcronym{UAV}{
  short=UAV,
  long=Unmanned Aerial Vehicle
}

\DeclareAcronym{DNN}{
  short=DNN,
  long=Deep Neural Networks
}

\DeclareAcronym{mmWave}{
  short=mmWave,
  long=millimeter-wave
}